\documentclass[12pt]{article}
\usepackage{graphicx}
\usepackage{amssymb,amsmath,amsfonts,palatino,amsthm}
\usepackage{amssymb}
\usepackage{epstopdf}
\usepackage{slashed}
\usepackage{color}
\newcommand{\f}{\begin{equation}}
\newcommand{\ff}{\end{equation}}
\newcommand{\blankline}{\vskip .3cm}
\usepackage{xcolor}

\begin{document}

\title{Realism and Causality II: Retrocausality in Energetic Causal Sets}

\author{
Eliahu Cohen${}^{1}$, Marina Cort\^{e}s${}^{2,3}$,
Avshalom C. Elitzur${}^{4,5}$
and Lee Smolin${}^{3}$
\\  \\
\small
${}^{1}$ Faculty of Engineering, Bar Ilan University, Ramat Gan 5290002, Israel
\\
\small
${}^{2}$ Instituto de Astrof\'{i}sica e Ci\^{e}ncias do Espa\c{c}o\\ \small
Faculdade de Ci\^encias, 1769-016 Lisboa, Portugal
\\
\small
${}^{3}$ Perimeter Institute for Theoretical Physics,
\\ \small 31 Caroline Street North,
Waterloo, Ontario N2J 2Y5, Canada
\\ \small
${}^{4}$ Iyar, The Israeli Institute for Advanced Research,\\
\small POB 651, Zichron Ya’akov 3095303, Israel 
\\ \small
${}^{5}$ Institute for Quantum Studies, Chapman University, Orange, CA 92866, USA}

\date{\today}
\maketitle
\begin{abstract}
We describe a new form of retrocausality, which is found in the behaviour of a class of causal set theories, called energetic causal sets (ECS). These are discrete sets of events, connected by causal relations. They have three orders: (1) a birth order, which is the order in which events are generated; this is a total order which is the true causal order,  (2) a dynamical partial order, which prescribes the flows of energy and momentum amongst events,  (3) an emergent causal order, which is defined by the geometry of an emergent Minkowski spacetime, in which the events of the causal sets are embedded. However, the embedding of the events in the emergent Minkowski spacetime may preserve neither the true causal order in (1), nor correspond completely with the microscopic partial order in (2).  We call this {\it disordered causality}, and we here demonstrate its occurrence in specific ECS models.

This is the second in a series of papers centered around the question: Should we accept violations of causality as a lesser price to pay in order to keep realist formulations of quantum theory? 
We begin to address this in the first paper \cite{CCES1}  and continue here by
giving an explicit example of an ECS model in the classical regime, in which causality is disordered.
\end{abstract}
\newpage
\tableofcontents


\section{Introduction}

The contemporary interest in retrocausality \cite{R1}-\cite{R12} is motivated primarily by the suggestion that it may be a route to a realist resolution of the paradoxes of quantum mechanics \cite{retroA1}-\cite{retroA5}.
In this paper we begin to investigate a new type of retrocausality which we have found in the classical regime of a type of dynamical systems -- energetic causal set models -- previously introduced by Cort\^{e}s and Smolin \cite{ECS1}-\cite{ECS4}. These models have a ``pre-spacetime'' similar to that of causal sets \cite{cs} with the difference that the current model allows exchanges of energy and momentum.

General relativity teaches us that most of the information carried by the geometry of spacetime encodes the causal structure, which is to say the causal relations amongst events.  This has led
to the following hypothesis concerning quantum gravity: {\it quantum spacetime consists
most fundamentally of a discrete set of events and their causal relations, and that the geometry of classical spacetime is an emergent and coarse grained description of bulk averages of those fundamental
causal relations} \cite{cs}. A number of models of a fundamental quantum causal 
structure have been proposed and studied, with the aim of demonstrating the hypothesized emergence of classical spacetime \cite{ECS1,cs,CDT,SF}. 

If this hypothesis is correct, the world has two causal structures: (1) the fundamental and microscopic causal structure, which presumably governs the Planck scale, and (2) an emergent, coarse-grained, and macroscopic causal structure, which appears at much larger scales, and at which a description of nature in terms of classical spacetime becomes possible.

It is then possible to ask whether the causal structures in the two regimes must always agree.
By this we mean whether the arrow from the past to the future defined 
by the underlying causal structure in (1) above, and by the emergent classical spacetime in (2) always align, and never contradict each other.  Most past work on the emergence of classical spacetime from models with fundamental causal structure assumes the micro and macro causal structures always agree.

The main result of this paper is to demonstrate that in the model of fundamental causal
structure which we developed in \cite{ECS1} (in which the emergence of a macroscopic spacetime has been 
shown)
the two causal structures often do not align.  These are
energetic causal set models, introduced and developed in \cite{ECS1}-\cite{ECS4}. Below we show by direct numerical simulation that the underlying causal order and the time direction in the emergent Minkowski manifold, sometimes agree and sometimes disagree.  

We call this phenomenon {\it disordered causality}, or {\it discausality.}  It provides the opportunity to study a specific example-in the classical regime-of the more general phenomenon of {\it retrocausality} referred to above, \cite{R1}-\cite{R12}.  These are models of fundamental physics in which the directions of causal influence within the different components of a  complex causal process can sometimes be opposed to each other.

The title of this work appears to violate an assertion made by us in the original proposal of energetic causal sets, namely that time and causality have a well-defined directionality that is never inverted, not even in principle.  
The results shown below challenge this assumption and ask whether the macroscopic arrow of time, as measured by macroscopic clocks, and the arrow of causality can ever be opposed. After detailed examination of the behaviour of the ECS model we will find in this study that the macroscopic arrow of emergent time can be inverted with regards to the arrow of causality between different events, but the order of causality itself is never inverted. This is a revision and clarification of the assertions made in our seminal ECS work.

Our model shows discausal behavior which is completely classical.  Yet, we believe it may be relevant for open questions in quantum theory\footnote{The possibility that energetic causal set (ECS) models might provide a completion of quantum mechanics was already proposed by us in \cite{ECS2, views}.}
 and quantum gravity because retrocausality has often been proposed as part of a resolution of the open problems in quantum foundations \cite{Wheeler}-\cite{Liar}. 
 A realist completion of quantum mechanics must necessarily account for the 
 non-local correlations that the experimental tests of the Bell inequalities tell us need to hold.
 One way to accomplish this is to combine processes that appear to go in opposite directions in Minkowski spacetime.  As a result, 
retrocausality \cite{R1}-\cite{R12} was proposed by a number of authors with diverse viewpoints as an elegant solution \cite{retroA1}-\cite{retroA5} to the problem of giving a detailed realist description of what goes on in each individual microscopic process. 

\blankline
\blankline

This is a companion paper to \cite{CCES1}.
It was a very interesting collaboration which resulted from an unexpected convergence of results in two very different lines of investigation regarding the foundations of physics. This convergence has spawned our collaboration, and resulted in these two papers. 

In the first paper, we examine several challenges to realism in quantum theory. Among other things, we investigated there the possible role of retrocausality in realistic approaches to quantum foundations.
More specifically, we explore there  the general proposal that retrocausality  may provide an answer to what exactly is going on in individual processes, which quantum mechanics only describes statistically. 
Thus, both papers in this series address the question: Should we accept subtle violations of causality as a small price to pay in order to keep realist formulations of quantum theory and understand the emergence of spacetime?





In the next section we review the energetic causal set models as a prelude to section 3, where we present results of numerical evolution of our model, showing the presence
of discausal processes and explain why they occur.  In section 4 we take a more detailed look at how the proportion of discausal processes depends on whether the model is in its disordered or ordered phase, and we relate these to the role that capture by limit cycles plays in the late term dynamics.  In a section titled simply, ``Becoming", we explain how retro- or dis-causal processes fit perfectly into a worldview in which the flow of time is the continual creation of novel events, out of an ever changing collection of present events.

\section{Review of energetic causal set models}


\noindent
ECS models were proposed in~\cite{ECS1} and~\cite{ECS2} motivated by the assertions: 

\begin{enumerate} 
\item the underlying laws of fundamental physics are time asymmetric, not time symmetric, as is the common belief; and 
\item causality is the fundamental principle governing all physical processes. 
\end{enumerate}

If we believe that quantum gravity is the fundamental regime, this implies that the laws of quantum gravity are not reversible in a time coordinate.
The goal of the program is to place the arrow of time as a main ingredient in the dynamics of the universe at all regimes.

\subsection{The basics of $ECS$ models}

ECS are causal sets \cite{cs} endowed by a flow of energy-momentum between causally related events.  
Energy and momentum are defined as fundamental in the ECS conception, and properties of events, while spacetime is emergent. We take the energy momentum conservation laws as fundamental properties of energy-momenta \footnote{
There have been proposals adding data to each event, subject to different principles, in the works of Ref.~\cite{fotini}}
Amongst the successes of the model so far, a new mechanism was introduced in the
aforementioned papers for the emergence of Minkowski  spacetime from pure causal connections in an energy momentum space. In particular we have shown the emergence of spacetime for the (1+1)d model. 
We have also examined higher dimensional models--where the dimension is controlled by the number of input and output events for each event, along exactly the same lines. We believe there is a good case for the extension to the 3+1 dimensional model, but we have not published the details of these constructions yet. Lastly the seminal paper also includes a route to generating emergent curved spacetimes, which has yet to be developed further.

In~\cite{ECS3} an identification was made between a spin foam model (constructed by Wieland in \cite{WW}) and ECS. This
made it possible to apply the new mechanism for the emergence of spacetime to the spin foam formalism.
In \cite{ECS4} we showed that these models also exhibit a transition from a time asymmetric phase to a phase of quasi-time symmetry, which has to do with the capture of deterministic dynamical systems by limit cycles.  To show this, we established a correspondence of ECS with a class of dynamical systems which similarly have an underlying irreversible evolution, but in the long term exhibit the properties of a
seemingly \footnote{In the sense that a globally irreversible dynamical system truncated to its limit cycles is reversible, because every state in a cycle has  a unique  antecedent as well as a unique descendent.} time reversible system in the form of limit cycles.

We therefore proposed that nature is ultimately described by a time-irreversible law, from which emerges a time-symmetric effective theory which governs phenomena at late times and large
scales \footnote{This was in fact proposed a long time ago by Penrose \cite{Roger}.}.
But if this is the case, then general relativity must be a late time limit of a time asymmetric theory,
which gives an effective description of the transition from irreversible to effectively reversible
dynamics.  We looked then for ways to extend general relativity to a time asymmetric theory, and
we found two.

In~\cite{TA1}, we introduced a new class of gravity models that extends general relativity by introducing a term proportional to the momentum, which therefore breaks time-reversal symmetry.  Ref.~\cite{TA2} compared predictions of the models thus derived to cosmological constraints available currently.  Then in \cite{GLdot} we found a time-asymmetric extension of general relativity in which both Newton's constant and the cosmological constant become evolving, conjugate degrees of freedom.

\subsection{The dynamics of  $ECS$ models}

We begin by reviewing the dynamics of the ECS models defined in \cite{ECS1}.

A causal set \cite{cs} is a set of events, connected by relations which are called causal relations.
An energetic causal set model describes a discrete or quantum causal structure, which grows by the sequential addition of new events, according to a simple set of rules. For the purposed of this study we are working with the {\it classical} version of ECS models, in which events are discrete, and not the corresponding quantum models.  
As such, events are labeled, $I,J,K,\ldots = 1,2,3, \ldots$ according to their birth order.
Each event is endowed with an energy-momentum vector, $p^I_a$, which lives in a momentum space,  $\cal P$.  We also assign to each causal link, $\langle IJ \rangle$,
connecting a parent to a child,
a pair of momenta, denoting the outgoing energy momentum, called $p_{a K}^I$, from event $K$ going towards event $I$,  and the incoming momentum, received by $I$ from $K$,
denoted, $q_{a I}^L$.  The difference between them, parameterized by  a parallel transport
matrix, which we call {\it the redshift}, is where the spacetime curvature may be coded.
It is set to zero by a constraint in the models we have studied so far.

A single ECS spacetime history is then represented by a series of causal sets, where each event and causal link are labeled by energy-momentum vectors, and each one is generated from the last by an action of what we call the ``event generator".    The event generator
acts at each step, $STEP_I$, to create a novel event, $E_I = I$ from two existing events, which can be said to be the parents of the new event.
Each created event has
a fixed number of ``children", we will choose $2$ here.  Each event also has a fixed number of parents which we will again take to be two \footnote{
Note that the ECS procedure for generating causal histories is very different from the classical sequential growth that generic causal sets make use of. In ECS we build a new causal structure by taking a new event and picking two in the present set via the procedure of extremizing a given quantity. In classical sequential growth two events are picked randomly and connected to a new event. In addition, in classical sequential growth there isn't a distinguished notion of present versus past, in the sense that the generation of new events does not happen as time progresses, as it does in energetic causal sets.}.

At each step, $STEP_I$, we may distinguish those events which have $2$ children, and call their union {\it the Past}, because they can no longer have a direct influence on the growing future.

Events which are still capable of being parents to new events, because they have one or no children, we call {\it the Present}.  We note that the  present is thick in the sense that two events
in the present may be causally related.

The future does not yet exist.

At each step, the event generator performs an optimization over all pairs of members of the present to choose the two that will be parents to the next event.  The optimization is over a measure of how  distinct the pasts of the pair's members are from each other.   Once a pair of parents is chosen, energy and momentum are distributed from the parents to the children, so that certain constraints are respected.  These are

\begin{itemize}

\item{}Conservation laws:
\f
{\cal P}_a^I = \sum_K p_{a K}^I -  \sum_L q_{a I}^L  =0,
\ff
where the sum over $K$ is taken over all events $I$ is connected to in the past and the sum over $L$ is over all events $I$ is connected to in the future.

\item{} No redshifts
\f
{\cal R}_{aI}^K = p_{aI}^K - q_{aI}^K =0.
\ff

\item{}Energy momentum relations for massless particles
\f
{\cal C}^I_K = \frac{1}{2} \eta^{ab} p_{a K}^I p_{b K}^I  =  0 , \ \ \ \ \  \tilde{\cal C}^I_K = \frac{1}{2} \eta^{ab} q_{a K}^I q_{b K}^I  =  0.
\label{em}
\ff

\end{itemize}

These may be expressed by a totally constrained action, which is extremized at each step to determine the energy momentum transmitted from the parents to the children; these are attached to causal links labeled by both the parent and the child.  The action is built by introducing Lagrange multipliers to express the constraints.

 \f
 S^0= \sum_I z^a_I {\cal P}_a^I   +\sum_{(I,K)} (   x^{a I}_K {\cal R}_{aI}^K + {\cal N}^K_I {\cal C}^I_K -  \tilde{\cal N}^K_I  \tilde {\cal C}^I_K ),
 \label{S0}
 \ff
where the sum over $(I,K)$ is taken over all connected pairs of events.

The Lagrange multipliers $z^a_I$ label points in a space dual to the momentum space $\cal P$.
In the simplest case, momentum space may be chosen to be a flat $n$ dimensional manifold with a Minkowski metric $\eta^{ab}$.  
In that case its dual space, $\cal M$, inherits the metric of momentum space: $\cal M$ is an $n$ dimensional Minkowski space with metric $\eta_{ab}$.  This may be considered an emergent description of the causal set spacetime into which the events are represented by the points $z^a_I$.  These are found by varying the action by the energy momentum vectors incoming and outgoing on each causal link.

\f
\frac{\delta S^0}{\delta p_{a K}^I }= z^a_I + x^{a K}_I + {\cal N} p^{a I}_K =0
\ff
\f
\frac{\delta S^0}{\delta q_{a  I }^K } = -z^a_K + x^{aK}_I -  \tilde{\cal N} q^{a K}_I =0.
\ff
Adding these two equations and using $ {\cal R}_I^K =0 $ we find
\f
 z^a_I -  z^a_K =  p^{a I}_K (\tilde{\cal N}_I^K  - {\cal N}_I^K ).
 \label{emergence}
\ff
The Lagrange multiplier $z^a_I$ then represents the
event $I$.  $ z^a_I -  z^a_K $ is then a space-time interval between event $K$ and event $I$.  It is a light-like interval proportional to
the momentum $p^{a I}_K$ connecting $K$ to $I$.  The constant of proportionality involves the Lagrange multipliers  $\tilde{\cal N} - {\cal N} $ which is consistent with the fact that the affine parameter along a null ray is arbitrary.

We choose the solution so that, locally in $\cal M$, in the causal structure defined by
$\eta^{ab}$,  $ z^a_I -  z^a_K$ is future pointing whenever $I$ is the child of $K$.

We note that the assignment of points and null rays of $\cal M$ to events and their causal relations are local.  To complete the definition of the emergent spacetime $\cal M$ must be assigned a time orientation and a global structure.  In the models we have studied, a periodic spatial
identification is imposed, making $\cal M$ a cylinder.  The identification may involve a shift along a time like direction, $\Delta t$, and we note that this affects the  causal structure of
$\cal M$, but not the birth order or the dynamically generated intrinsic partial order, connecting children to parents, and governing the flow of energy-momentum.  This is one reason that
these two partial orders may differ.

\subsection{The basic phenomenology of $ECS$ models}

In order to test the assertion that the fundamental laws could be time-asymmetric we began by
exhibiting a class of dynamical systems which are time-irreversible, but from which emerges
relativistic particle dynamics, which is reversible.  These are the ECS models.
This is a large class of models, but we focused here on the same model analysed  in the seminal work which is a $1+1$ dimensional model studied extensively by with numerical simulations for a large variety of parameters and initial conditions.

\begin{figure*}[h!]
\begin{center}
\includegraphics[width=0.8 \textwidth]{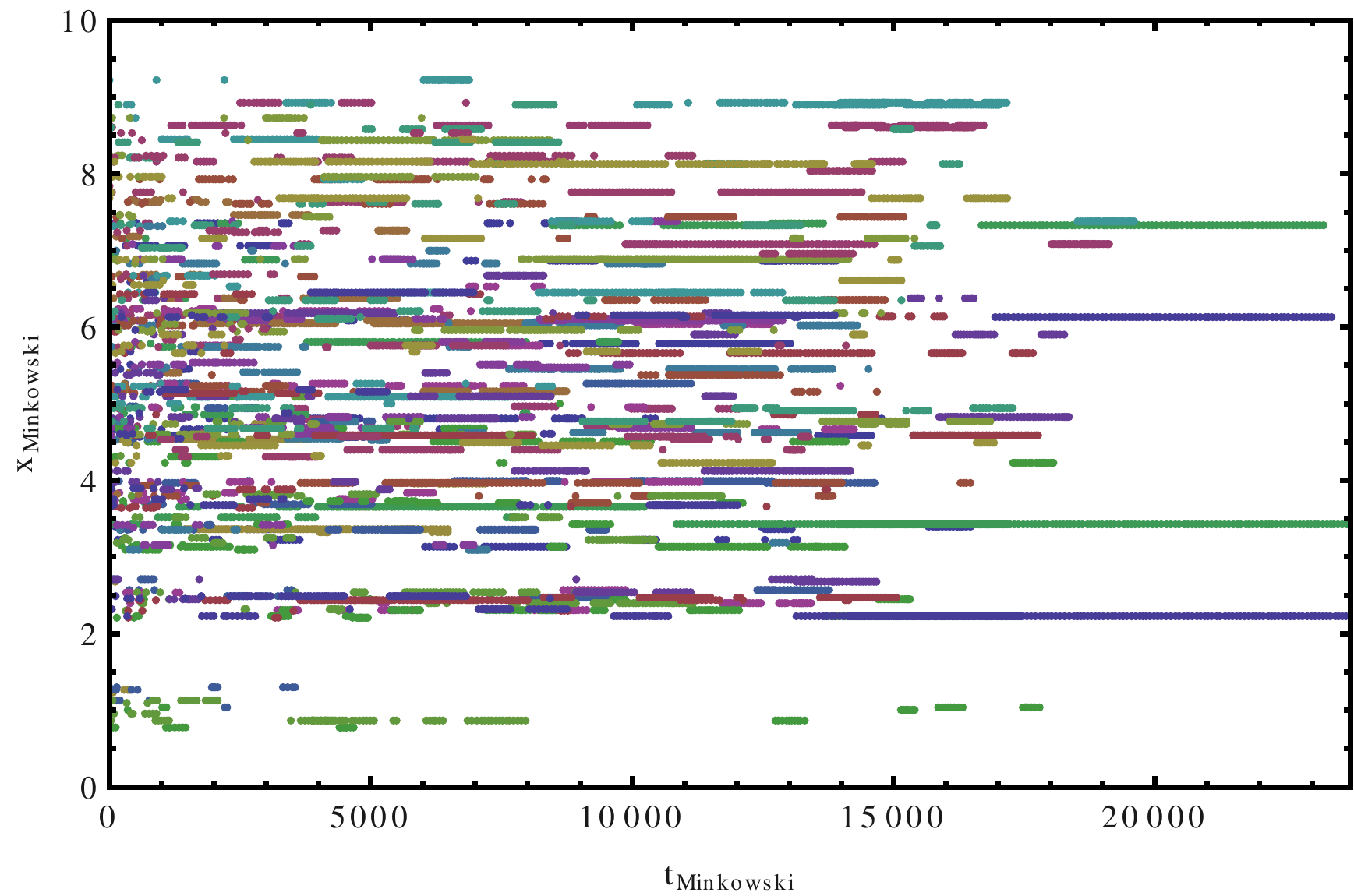} 
\caption{Simulation of a typical energetic causal set model, with (1+1) dimensions and a cyclic space coordinate. This model has 20 different families and total event number $10^4$. The figure depicts events as single dots in the emergent (1+1)d Minkowski spacetime. For the purposes of the current work the main aspect to highlight in the simulation is the emergence, at late times, of regular lines which we call quasi-particle trajectories. This marks the transition of the time irreversible to the time reversible phase.
\label{space_time_full}}
\end{center}
\end{figure*}

Here we show an example of such a model in Figure~\ref{space_time_full}. This model evolves under the time-asymmetric laws proposed.

Figure~\ref{space_time_full} depicts events generated in our ECS models, and in the Minkowski manifold. Note, that this is the classical version of the model proposed in~\cite{ECS1} and not the quantum version discussed in  \cite{ECS2}. We have shown that this spacetime emerges from the underlying causal set network of events which live in energy-momentum space.

The algorithm used to generate new events in the causal structure in Figure~\ref{space_time_full} is the same that was proposed in the original paper of the energetic causal set program \cite{ECS1}, and consists of extremizing a quantity associated with each event in the current (thick) present.
Each dot in the plot represents an event in the causal set, and each event marks the intersection of light cones of particles which live in energy-momentum space alone.
Different colors denote different families of ancestry as per the usual ECS model. The number of families denotes the number of degrees of freedom in the initial conditions. This is simply how many distinct elements there are at $t=0$. These elements will interact, create new events, and generate their own family. New events get stored in the family of one of the progenitors. Events belonging to the same family all share a common causal past, and all have the same color. 
The model analyzed for the purposes of the current work has 20 families -- though the choice of this number depends only on computer capacity and does not qualitatively alter results. If the number of families increases the only difference is a corresponding increase in length of time, i.e. the number of total events, that it takes for the system to coalesce to the symmetric phase. As we describe below, the transition to the symmetric phase is signaled by the emergence of quasi-particle trajectories.
Time runs in the horizontal direction, with events to the right taking place later. The vertical axis represents the 1-dimensional space coordinate which is cyclic, so the full (1+1)d manifold forms a cylinder with the time coordinate always increasing.

In figure~\ref{space_time_full} and in the simulations of \cite{ECS1} we saw evidence for a simple characterization of the time evolution of the causal networks in two phases. The system begins in a disordered phase, followed by an ordered phase which we call the locked-in phase. This two-phase structure was observed in runs with a wide variety of choices for the algorithm generating the events.
Each causal set begins with a period of apparently chaotic behaviour, embodied by structureless and disordered spacetime positions, which reveal the time asymmetry of the algorithm. In this early phase, the events form a roughly random pattern in space-time, showing a large uncorrelated display of their spatial positions. The pattern in this phase is asymmetric under time reversal. 
Then, as time progresses, we gradually begin to see brief trajectories, which are momentarily stable. These trajectories reveal an underlying locking between pairs of families in the causal network, and last longer each time they occur. 
Eventually they give rise to recognizable trajectories in the Minkowsky spacetime embedding, which we call trajectories of so-called quasi-particles. In this second phase, the locked-in phase, the patterns formed by the events are invariant under time reversal.

The emergence of stable trajectories of quasi-particles marks the transition from time asymmetry to time symmetry. These results show by example that an underlying rule which is asymmetric in time may evolve and coarse grain to exhibit apparent time symmetric evolution. This is by no means a trivial result given that the rule is time asymmetric throughout the whole evolution of the causal set.

\section{Retrocausality in ECS: disordered causality}

More recently,  the authors of~\cite{ECS1} stumbled upon the possibility that their models exhibit a form of the retrocausality that Cohen and Elitzur and others had discussed \cite{R1}-\cite{R12}.
This occurred while working on the ECS simulations for a different purpose, namely the identification of the ECS models with discrete limit cycles in \cite{ECS4}. During that study we discovered by accident that the underlying causal order of events-which is the order in which they are generated,
 does not always align with the time ordering of the emergent spacetime. This raised the suspicion that the causal order of the ECS structure does not always reflect the macroscopic arrow of time. We wished therefore to explore this effect further by reaching out to the community that studies retrocausal effects in foundational quantum mechanics.


\subsection{The three different orders and partial orders}

For the purposes of this paper, the key point is that there are 3 kinds of causal relations in the model, each
dynamically generated, and they need not always agree.



\begin{enumerate}

\item{}Birth order, which is the global and total order of event creation.

The model generates a sequence of events, $E_1, E_2, \dots $ one at a time by means of an {\it event generator}.  This gives us first of all, a total ordering of the set of events,
which we call the birth order.  Note that in choosing which are the events that are the parents to the next event, the event generator computes an optimization over all possible pairs of present events.
This process is global, hence the birth order is a genuine causal order.

The birth order is denoted
\f
 F \gg E   \ \ \  \mbox{for} \ F \ \ \mbox{is born later than} \  E.
\ff


\item{}Descendent ordering, i.e. ordering through chains of descent from parents to children.  This is a partial order.  We denote it by
\f
 F > E,  \ \
 \ff
which is true when there is a chain of descent that starts with $E$ and goes up to $F$,
i.e. $E$ denotes a grandparent or parent, or some more distant ancestor,
and $F$ is their child.

We noted above that at each step, the events which have so far been generated, $\{ E_1, E_2, \dots , E_I \} $
at step $I$, are divided into a present set and a past set.


At step $I$, the event generator performs an optimization among all the then-present events, and chooses a specific number, $N_{parents}$ of them to be parents to the next event.

To review, each new event that is created $E_{I+1}$ is deemed to be causally in the future of its parents.
Thus if there are three parents, $E_{1024}, E_{171717},   E_{171719}$ we have three new causal relations.
\f
E_{I+1} > E_{1024};  \ \ \ \ \ \ E_{I+1} > E_{171717}, \ \ \ \ \  E_{I+1} >E_{171719}.
\ff
We call this the dynamical causal order.  Note that this is a partial order, and that it is related
to the total birth order by
\f
E_K > E_J   \rightarrow  J < K,
\ff
but the converse is not necessarily true.  It is not the case that $E_K > E_J$ for all $J < K.$

Note also that the present is thick, that is two events can be part of the present at step $I$,
but be causally related.

The dynamics also distributes the energy and momentum of each event to its children.  Thus the dynamical causal order is the partial order that the flow of energy and momentum respect.

Let us now consider two adjacent events in birth order:  $E_J$ and $E_{J+1}$.
We have said that it may be the case that in the causal order
$E_{J+1}  > E_J$.  But it is also possible that $E_J$ and $E_{J+1}$ are causally unrelated
in the dynamical causal order.

Now we come to the third ordering, which has to do with the mechanism by which there emerges
a Minkowski spacetime, $M$, such that for each event $E$ there is a point $z_E \in M$.

\item{} Causal ordering in the embedding of the events as points/events in the emergent Minkowski spacetime, $M$.

In \cite{ECS1} a procedure is given for embedding the events
of the model to points in a Minkowski spacetime.  This is developed
in detail in a $1+1$ dimensional model, with $N_{children}=N_{parents}=2$, and it is found to be always possible.  The image of event $E_J$  under the embedding, is the point $z^a_J$ of
Minkowski spacetime.

This is also a partial order, which we denote by
$z_F >_M  z_E$.  This   means that the point $z_F \in M$ that represents the event $F$ is to
the causal future in  $M$ of the point $z_E$ that represents the event $E$.
\end{enumerate}

It is important to note that the solutions to (\ref{emergence}) and hence the emergent causal structure, depends
on the global structure of $\cal M$ as well as its conformal metric $\eta_{ab}$.

A partial diagnostic of the Minkowski causal relation, is the Minkowki time coordinate,
$t=z^0$, of $\cal M$, in the sense that $z_F >_M  z_E$ implies
that $z^0_F >  z^0_E$ (but, of course, not always the reverse.)  We note that this is unique as
the cylindrical boundary conditions break Lorentz invariance.

\subsection{How disagreement can arise among the three orders}

It is easy to see that we have the following relations amongst the three orders.

\begin{enumerate}

\item{}

    $E_K > E_J   \rightarrow z_K >_M z_J,
    \ \ \ \     \mbox{ but the reverse need not hold.}$
    
\item{}

$\mbox{If}  \ F>E  \ then \ F \gg E,    \ \ \ \     \mbox{ but the reverse need not hold.}$


\item{}Hence, it is possible that  $z_F <_M z_E$, while at the same time   $F$ and  $E$ are unrelated under the causal order.

\item{}Because the present is thick it may happen that $J \gg K$ while $z_K >_M z_J $.

\item{}As a consequence, increases in birth order may not always be aligned with increases in the Minkowski time coordinate.  Some times a later-born event may be represented
by an earlier $t=z^0$.

\item{} Even if $I>J$, it can happen in some models that
$z_J >_M z_I$.
(One way to generate examples of this is to chose the shift $\Delta t \neq 0$ large
enough to allow closed causal loops under $<_M$.)

\end{enumerate}



When {\bf (4), (5)} or {\bf (6)}  occurs we say that {\it causality has been disordered.}

\subsection{Disordered causality appears in $ECS$ models}

In order to test this possibility, we superposed the causal links of the network of events on top of the dots representing those same events in the emergent spacetime.
That is, we keep the events in the emergent Minkowski spacetime, but connect these by their causal links (in blue and red), which connect the events in the order that they were generated. These links make up the causal network, and which are usually hidden in the Minkowski spacetime representation.
The model represented is the same as in Figure~\ref{space_time_full} but the dots representing the events have now been removed in order to allow for visualisation of the causal links. Therefore events are now at the two ends of the causal links connecting them.

\begin{figure*}[h!]
\begin{center}
\includegraphics[width=0.8\textwidth]{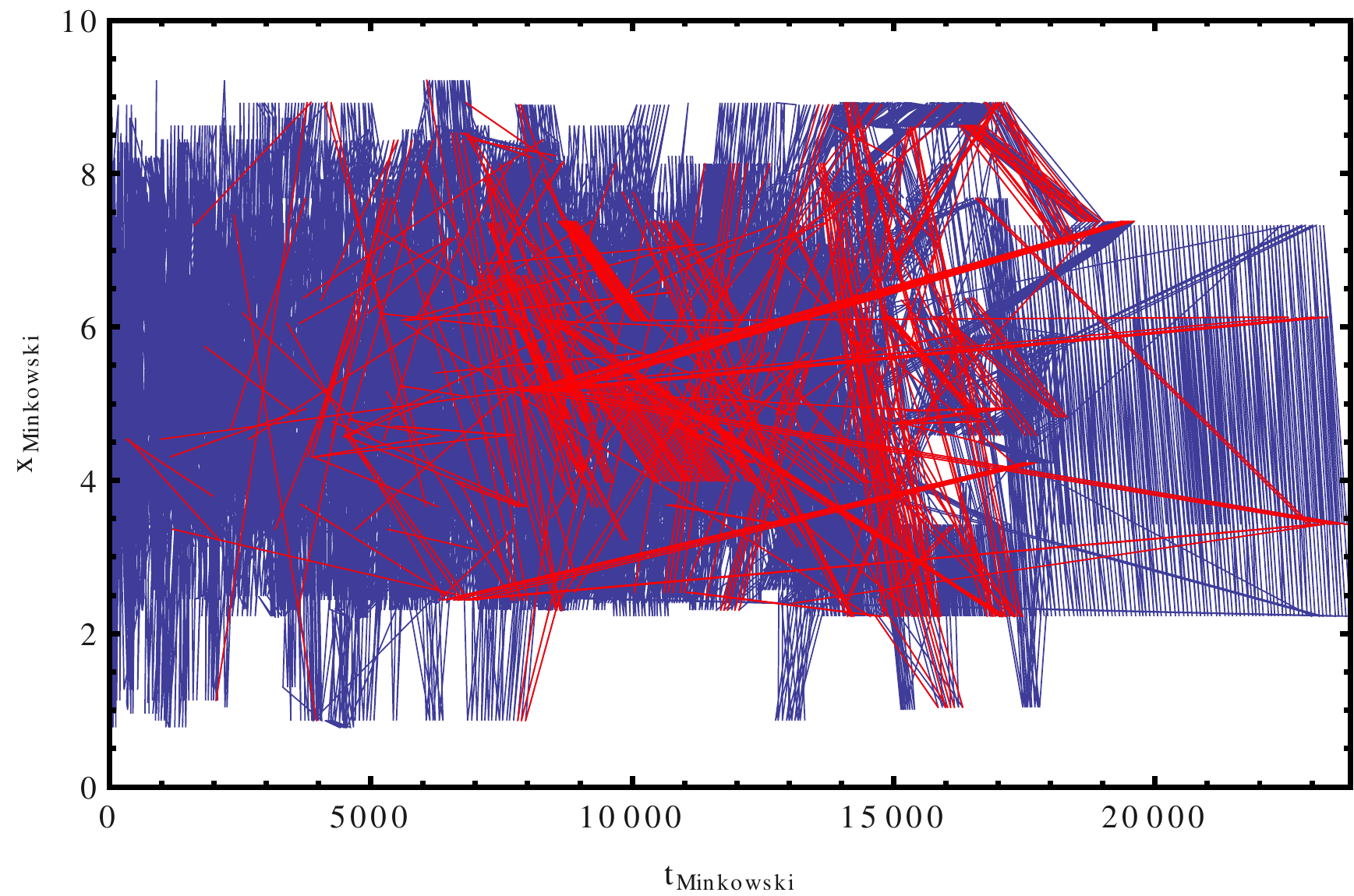} 
\caption{Events generated in a (1+1)d energetic causal set model with 20 different families and total duration of $10^4$ events. This panel shows the same events as in Fig.~\ref{space_time_full} but now with the underlying causal links of the network uncovered. That is, this figure links the events one by one in the order that they were generated. For comparison purposes with Fig.~\ref{space_time_full} we have left the scale of this plot unchanged, though this means density of causal links is high and the individual links are indistinguishable. In red we show the causal links which are highly discausal, that is events which move against the direction of time in Minkowski spacetime. We display in red only large discausal jumps, with a backward jump which is more than $1/20$ of the full time range. See text for explanation of this individual choice of 1/20-th jump size. In Figure~\ref{zoom_in} we will zoom-in and reveal distinct regions of the causal set for different stages of  evolution.
\label{eventsSequence}}
\end{center}
\end{figure*}

The result is in Figure~\ref{eventsSequence}.
The right panel shows the same Minkowski spacetime axes as Figure~\ref{space_time_full} , as well as the same set of events, but now they are linked by the order in which they are generated as the causal sets evolves.

\begin{figure*}[h!]
\begin{center}
$\begin{array}{cc}
\includegraphics[width=0.5 \textwidth]{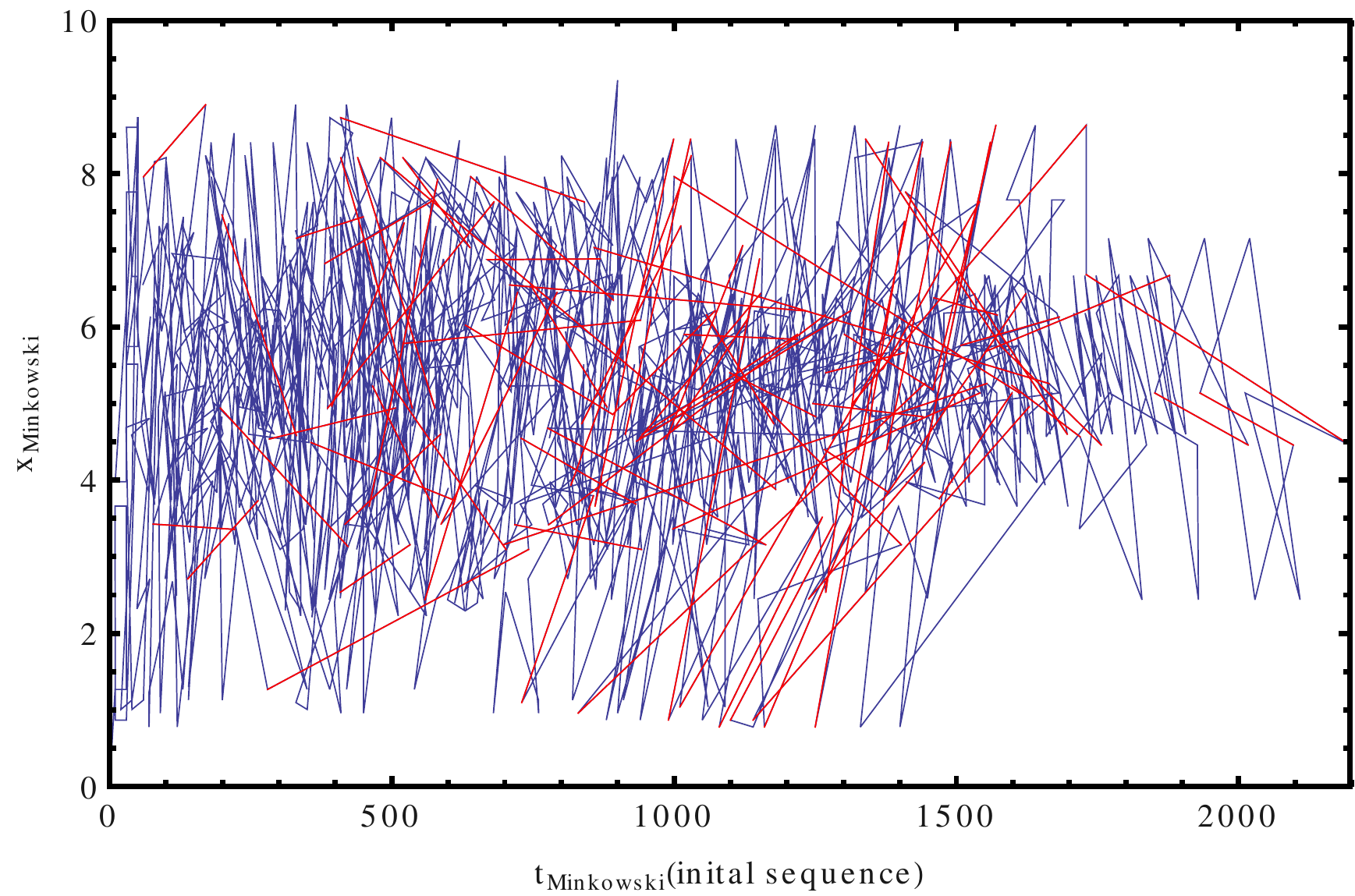}&
\includegraphics[width=0.5 \textwidth]{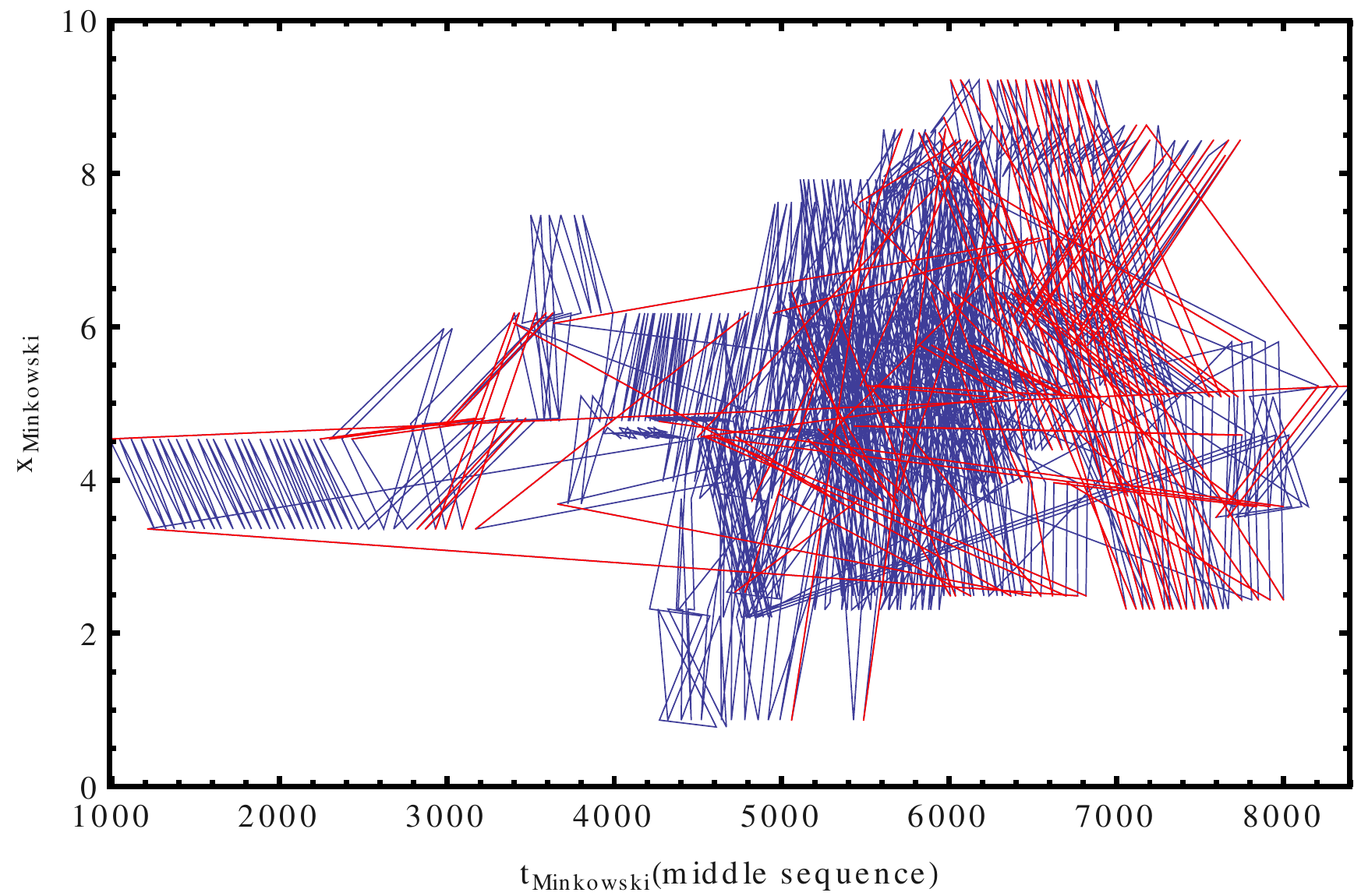}\\\\ 
(a) & (b)\\\\
\end{array}$
\includegraphics[width=0.5 \textwidth]{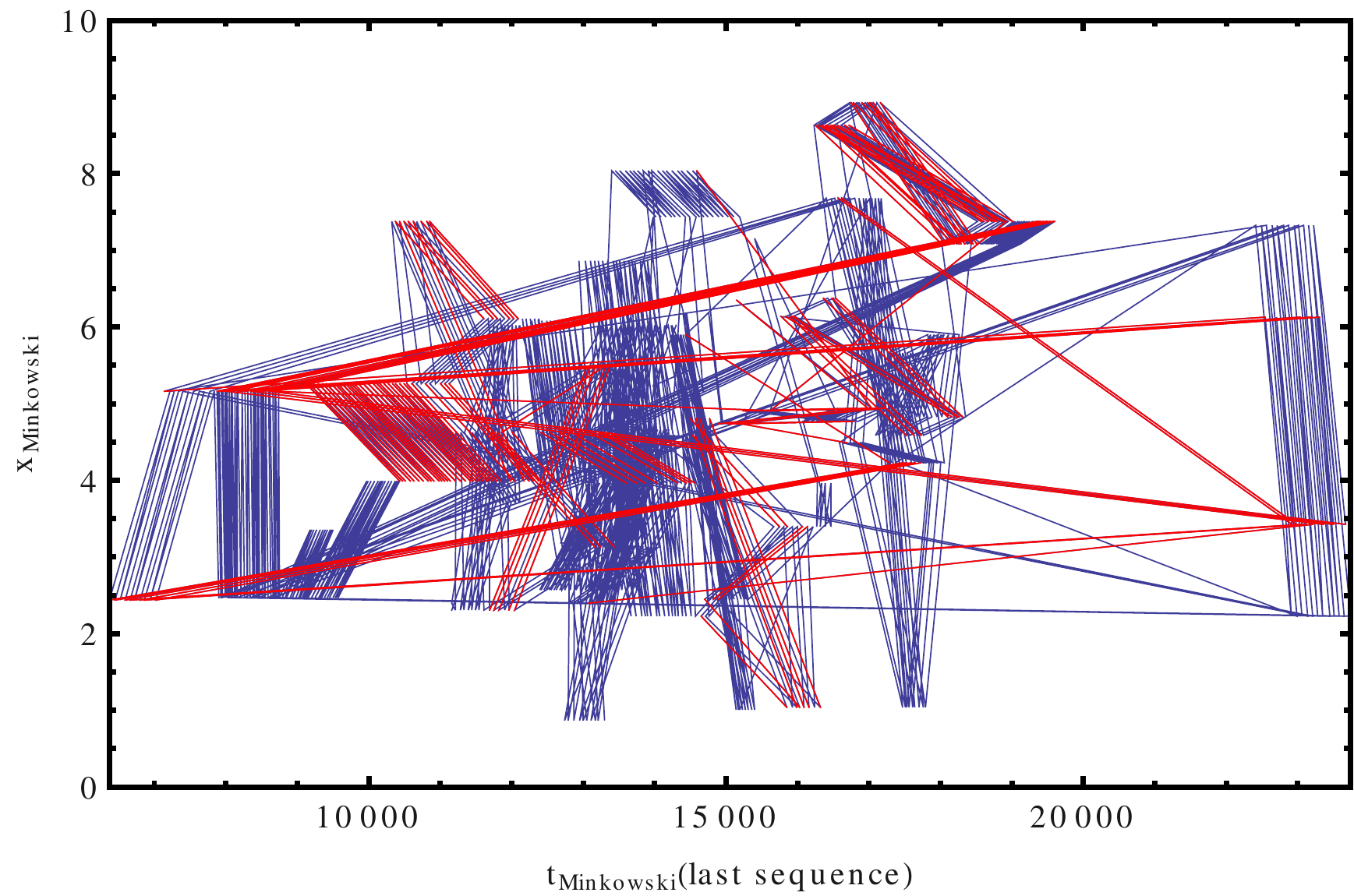}\\
(c)
\caption{Zoom in on the full sequence of the causal set structure, represented in the emergent Minkowski spacetime. The panels depict {\bf (a)} the first $10^3$ events, {\bf (b)} $10^3$ events in middle sequence, and {\bf (c)} the last $10^3$ events in the sequence. The blue lines represent the causal links. Red displays highly discausal jumps: moves that jump backwards in time with amplitude larger than $1/20$ of the time scale of each individual plot.
Although the amplitude of discausal jumps increases with the direction of emergent time, their frequency decreases as the causal set evolves. This is in compliance with the overall evolution towards time symmetry of the structure.
\label{zoom_in}}
\end{center}
\end{figure*}

This first result of Figure~\ref{eventsSequence} is not particularly illuminating because the density of causal links in a simulation with as many events as those of a typical causal set is very high.
For that reason in Figure~\ref{zoom_in} we show truncated portions of the full set evolution of Figure~\ref{eventsSequence}, zooming in on the high density of points there.

However Figure~\ref{eventsSequence} does suffice to suspect that the causal structure or birth order may not be trivially inferred from the ordering given by the lightcones of Minkowski spacetime. Hence the evolution of the causal structure and the Minkowski time may at times have opposing directions.



Instead, we can already see from the crowded display of Figure~\ref{eventsSequence}, that there are many cases where two subsequent events, in terms of birth order, are represented by
links that are space-like (represented by vertical links in the figure) or even time-like, but oriented to the past.

So from Figure~\ref{eventsSequence} we infer that the causal evolution does have clear instances of violating the Minkowski time direction and instead zigzaggs back and forth in the direction of time of the embedding.  This is  in clear opposition to the classical notion of causality which is aligned with the arrow of time.

\begin{figure*}[h!]
\begin{center}
\includegraphics[width=0.8 \textwidth]{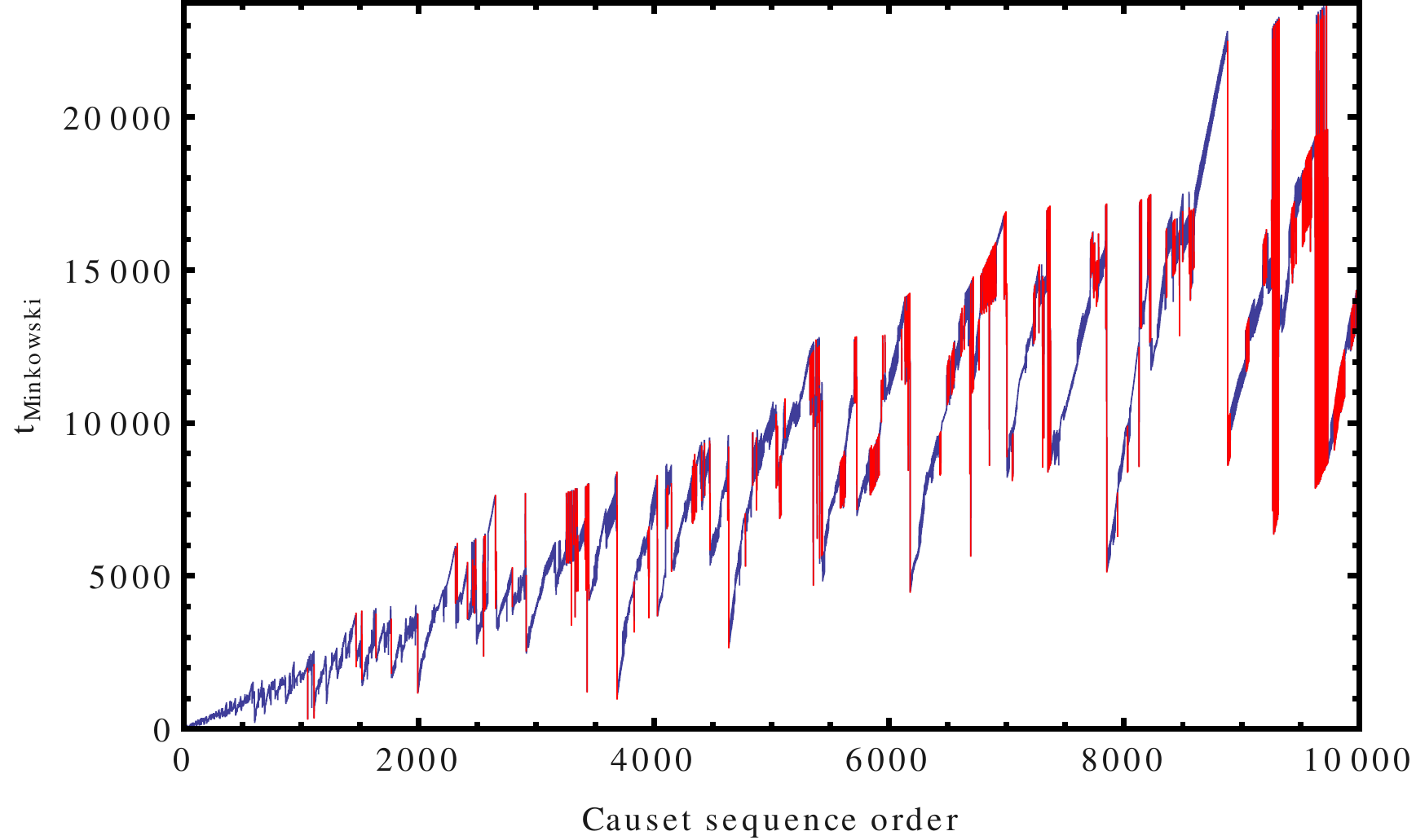}
\end{center}
\caption{Minkowski coordinate time versus birth order.  A causal set sequence is plotted in order to trace the frequency of discausal jumps in the sequence. If causal jumps were always aligned with emergent time the plot would be monotonically increasing. Downward moves represent causal jumps opposing the arrow of time. In red we highlight large discausal jumps: we choose these to be moves that jump backwards in time with amplitude larger than a threshold of $1/20$ of the full time range of the plot. See text for choice of threshold value.
\label{MinkTime}}
\end{figure*}

This becomes even more evident in Figure~\ref{MinkTime} where we plot the evolution of Minkowski time on y-axis, versus the event order in the causal set sequence in the x-axis. If the two directions of causality and of emergent time were always aligned, this plot would be a monotonically increasing function. Instead we see that as the events take place on the x-axis, the corresponding value for the time of the embedding retrocedes, sometimes by a large amount, to an earlier time of the embedding. In red we signal discausal jumps which are going backwards in time by an amount larger than 1/20 of the full Minkowski time scale. The need for a threshold for what we consider to be a retro- or discausal move is there because all moves in the causal set are either slightly forwards or backwards in manifold time. Since there are no moves of infinite speed allowed by the model, most of these moves which are backwards in time are just portraying the normal dynamics of the model, as dictated by the algorithm, and should not be interpreted as opposing the emergent arrow of time in the manifold, that is should not be interpreted as discausal.

Instead, the moves we are interested in -- the ones which we call discausal -- are moves which interrupt the dynamics of the event generator in a particular patch of space-time, and go back in time (by an amount large enough compared to the full time scale) to pick up events in the past that had been left behind by the dynamics. These are the discausal moves which are interested in, those which bring back events that remain in the present after many moves have been made, to participating in the generation of new events. The value of the threshold is optimized to distinguish between the two types of moves.
We assessed this amount to be roughly $1/20$ of the full time scale by observing that $1/10$ would exclude many moves which clearly oppose causality, and $1/30$ would include many moves who are part of the causal set dynamics and not discausal.

In order to examine what is at hand we zoom in on the causal sequence of Figure~\ref{eventsSequence} for better assessment of the opposition of the two evolutions. In the upper left we plot of Figure~\ref{zoom_in} we zoom in on the first 1000 events of Figure~\ref{eventsSequence} and show their causal ordering displayed in the same spacetime embedding.

On the upper right panel we zoom in on a sample of events in the middle of the causal set, taking events 4000 through 5000 (the full causal set is formed by $10^4$ events). Finally, in the bottom panel, we show the final 1000 events of the causal set.

In this figure we examine in detail how the generation of events is ordered compared to Minkowski time. In all three panels we see instances where the causal link first goes forward in the direction of Minkowski time but retrocedes afterwards.

All three panels depicting the initial, middle and end of the causal evolution are very different qualitatively.
In the first panel the large time-like jumps are very common, and structurally very disordered so most moves violate the time direction of the embedding.
In the panel depicting the middle of the evolution most jumps have settled in the form of limit cycles. This means that the moves in the causal set have been trapped in a cyclic dynamics and oscillate between a small number (often just two) of space positions, resembling reversible evolution. The finding that ECS models settle in pools of apparent reversible dynamics which resemble time symmetric evolution was the result of Ref.~\cite{ECS4}.

In the last panel most of the jumps are now oscillations around spatial positions in the form of limit cycles. At the end of the simulation limit cycles trap most of the causal structure and are a confirmation that the system is now in the full time symmetric form. We see from the last panel that the structure of the causal links is now very regular, most of the jumps are moves in a cycle of two or three spatial position, and the jumps which are out of sequence with Minkowski time, the retrocausal jumps are now very rare.

\section{Discussion}

What we observe from this analysis is an arrow of causality which \textit{zigzags} with respect to the evolution of the metric tensor of the Minkowski spacetime. This occurs because sometimes the algorithm creating the new events chooses parents $A$, and $B$, which are in past of the current event $J$ in terms of Minkowski time.

This is possible because $A$ and $B$, despite being in the Minkowski past are both in the thick present, defined as the set of events still available to create new events. They are in the thick present with respect to their causal order but their coordinates in the Minkowski embedding, $z_A$ and $z_B$, are in the past of $z_J$, the embedding of the current event (last event created). As a result, it can happen that the next event, $J+1$, daughter of events $A$ and $B$ has a Minkowski representation, $z_{J+1}$, that is in the Minkowski past of $z_J$. That is $z_{J+1}$ in the past with regards to embedded spacetime, even if the causal oder is always respected. Also, as is a common result of generic causal sets, it may also happen that $J$ and $J+1$ are causally
unrelated in the dynamical causal structure, $>$,  given by ancestry.

This zigzagging of the action of the event generator in the emergent Minkowski spacetime forms what we call a \textit{causal corridor} in the manifold.


At this point we may ask a few questions.

First of all, might this disordered causality evolution allow for closed causal loops (which are forbidden by general relativity)?
The answer is no because the order of causality, as measured by both birth order and causal structure, is never violated.  A causal link once laid down, has a well defined direction that will never be violated (traversed backwards). Causal links may never be travelled through in the opposite direction. This also answers the question of whether the causal set may allow for closed time-like loops, in which events come to be affected by their image in the past. The answer is also no: once events have given origin to their full offspring, they are out of the thick present, i.e. they are part of in the past set, and will not again enter the algorithm for generating new events\footnote{The causal structure of an ECS encodes the fundamental rule that an event has only a finite number of descendants. Once those have been exhausted (have been generated) that event is archived, it is no longer available for interaction, and will not feature in the causal structure again.}.

In summary, in the ECS sense retrocausality does not refer to a violation of causality, but is instead a misalignment of the arrows of causality and the arrow of time in the emergent manifold.

So we have now identified two very different types  of violated causality

\begin{enumerate}
\item Strong violations of causality, which will involve causal loops.  We can also call this time symmetric retrocausality, as it tries to restore the time symmetry that is broken by the irreversibility of the usual quantum measurement postulate.

\item Weak violation of causality, which is the phenomena we discovered is happening in the ECS models. Here, causality is still fundamental, irreversible and time asymmetric, but it can sometimes run against the direction of time in the macroscopic and emergent spacetime. In order to distinguish it from the pure retrocausal sense, we refer to this phenomenon as disordered causality, or ``discausality''.
\end{enumerate}

\subsection{Discausality and the transition to the symmetric phase}

We have seen in earlier papers \cite{ECS1} and \cite{ECS4}, that these ECS models evolve through two phases, a first, chaotic time asymmetric phase and a second, ordered, apparently time symmetric phase.  It is then interesting to ask how the amount of discausality,  which may be taken to be measured by the proportion of discausal moves, changes with the transition between these two phases. We want to look at this question to assess whether there is a correlation or anti-correlation between the amount of discausal structure in the causal set and the time reversibility of the dynamics.

Even a cursory glance at Figure~\ref{zoom_in} reveals stark differences of the scale of jumps between all three panels. This signals an evolution in the amount of moves which oppose causality, occurring as the systems transits from the irregular asymmetric phase towards the organized sequence of the time symmetry stage.

We recall that the images of the events trace different kinds of patterns in the two phases:
first chaotic and time asymmetric, later ordered and apparently time symmetric.   Similarly, looking at the causal connections between events,  we see the level of organization in the structure of the set increase, as we progress from the first through to the third panel, from early to late times.

\begin{enumerate}

\item In the first panel the causal set is mostly disordered and is strongly in the time asymmetric phase. In this panel the alignment of the arrows of causality and of emergent time arrow, signalled by vertical or almost vertical jumps, is almost inexistent. At this early stage the simulation shows jumps mostly at random in Minkowski space.

\item In the second panel we begin to distinguish the organization of new events in pairs of two alternating space positions. This signals the transition of the system to the ordered phase,
dominated by  limit cycles, most of which are composed of two events. We studied this phenomenon in \cite{ECS4} and identified that the limit cycle phase corresponds to the emergence of the time symmetric regime, as well as the emergence of the quasi-particle trajectories.

\item In the last phase of evolution, in the third panel, the system has fully transited to its limit cycle phase. It now spends most of the time in limit cycles, which for ECS take on the form of a dialogue between two quasi-particles with events occurring alternately between their two spatial positions. At
irregular intervals the system exits a particular cycle in order to be caught by another. Every time this happens -- every time it switches between two limit cycles -- there's a large jump in the image of the process in the emergent Minkowski space.  This is often associated with
a large jump between the images of the events, marked by a long line going forward or backward in Minkowski time. In the context of this work this can signal a retrocausal jump.

\end{enumerate}


We established in previous works that organization in spacetime corresponds to a time symmetric phase, in which the dynamics are reversible with respect to $t$. Likewise disorganization, or irregular clustering of the events in the emergent manifold, corresponds to time irreversible evolution. Furthermore, as a follow-up of this result, in Ref.~\cite{ECS4} we established that organization in the emergent spacetime signals the occurrence of limit cycles: the system starts off in the strong time asymetric phase, as dictated by the evolution rule of the causal set, and progressively more and more events get caught in a limit cycle. When the system is in its strong time symmetric phase it is fully trapped by limit cycles.

The novelty of the current work lies in us having found that the time asymmetric phase corresponds to a phase when the system behaves often against the direction of causality. The rate of discausal jumps in the disorganized or irregular phase is very high. We see from Figure~\ref{zoom_in} that the rate of discausal jumps evolves from the first to the third panel.\\

\textit{In particular we assert that in the strong time asymmetric phase the moves in the causal set are highly discausal, and later, as the system becomes time symmetric, the moves of the causal set are mostly aligned with the direction of time, and the occurrence of discausal jumps is rare.}


\subsection{Discausality and system capture by limit cycles}
In order to test the hypothesis that the amount of discausality is anti-correlated with the evolution to time symmetry, we can trace the number of moves that the system spends in limit cycles, and plot its evolution.

\begin{figure*}[h!]
\begin{center}
$\begin{array}{cc}
\includegraphics[width=0.5 \textwidth]{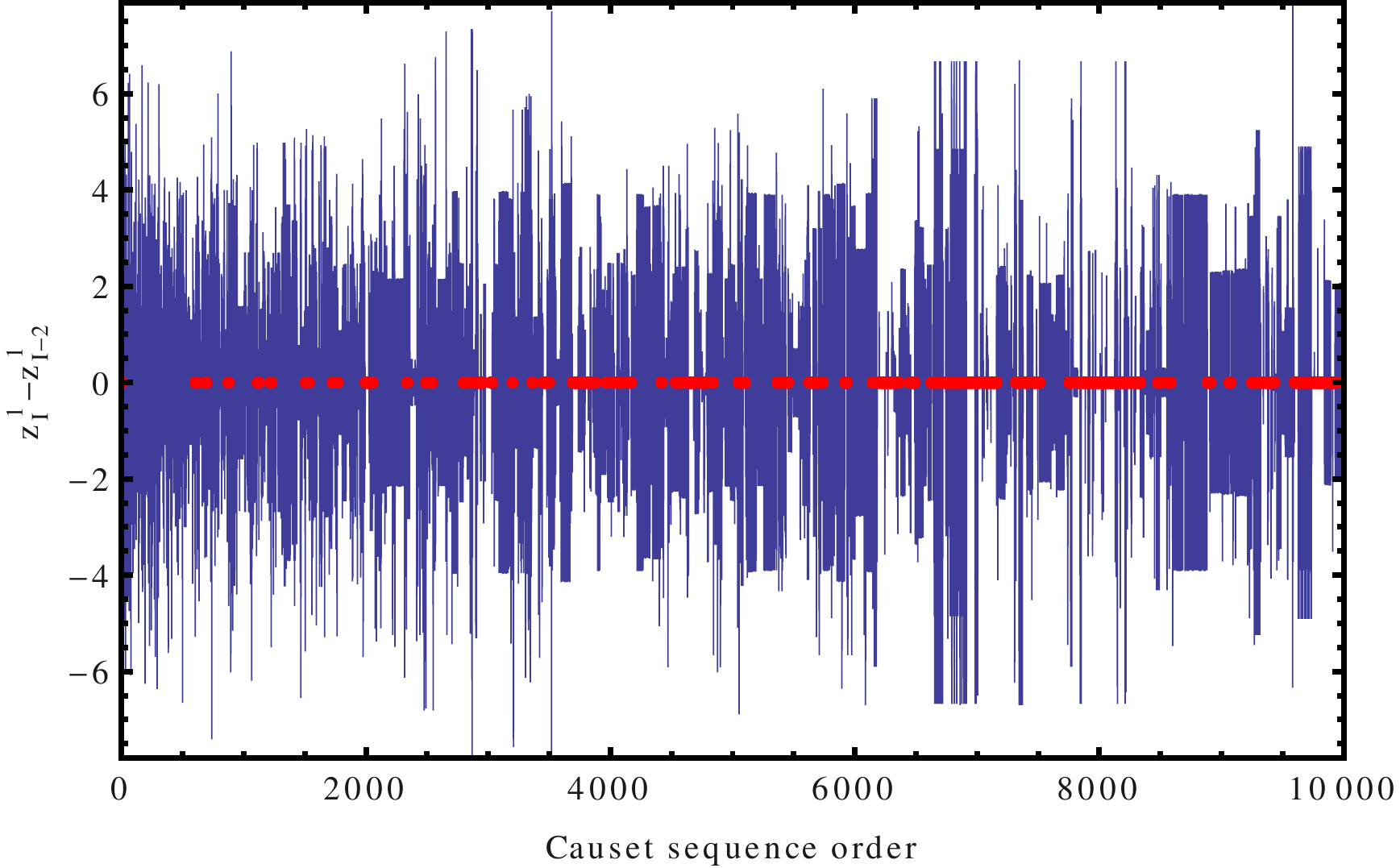}&
\includegraphics[width=0.5 \textwidth]{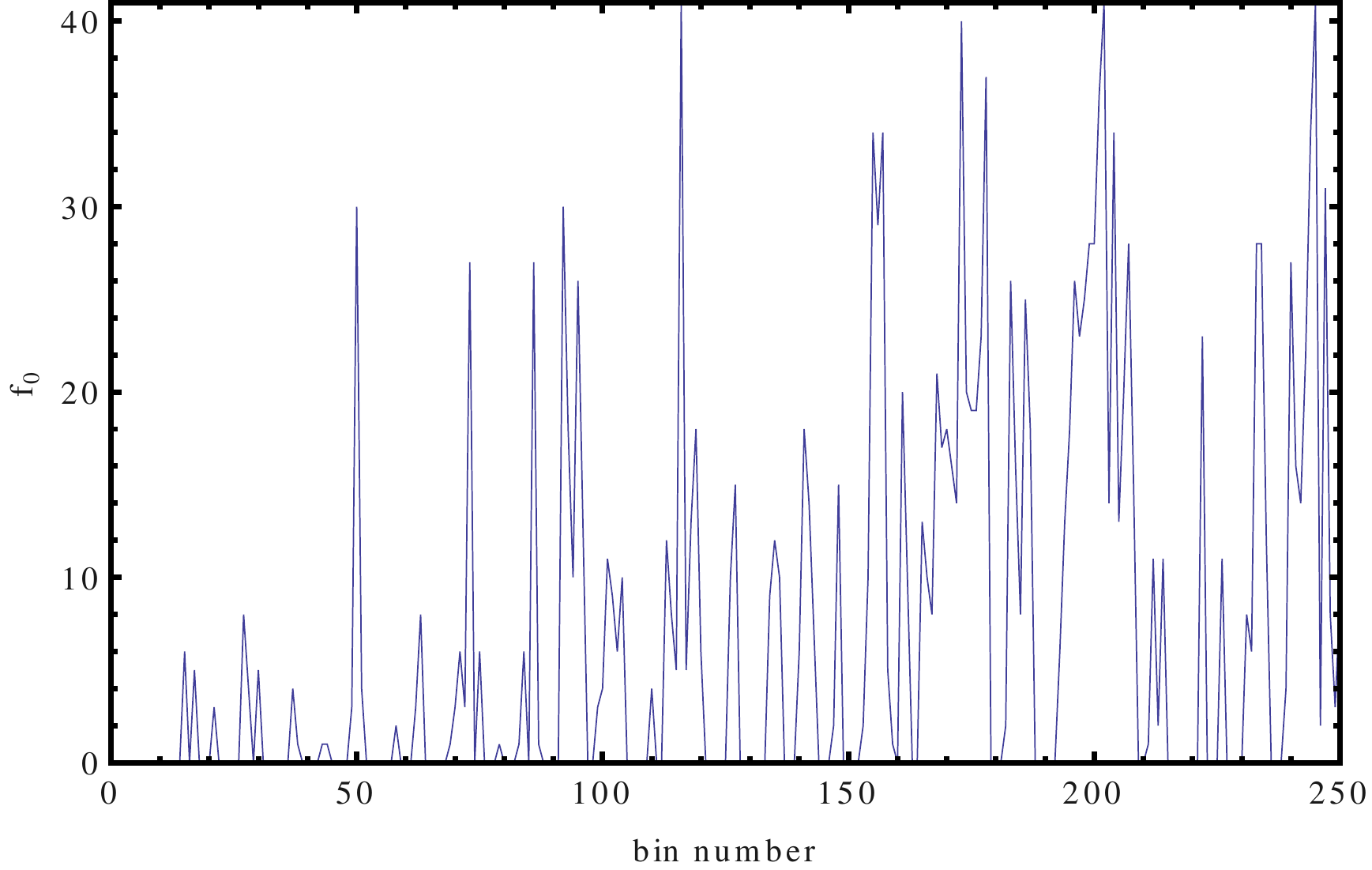}\\\\
(a)&(b)
\end{array}$
\caption{\textbf{(a)} Difference between consecutive pairs in the causal set sequence plotted to investigate which jumps are retrocausal. A zero value signals moves which are causally ordered and a non zero value signals retrocausal moves. The frequency of occurring zeros in this plot indicates then that the arrows of causality and emergent Minkowski time are aligned. The system starts off strongly discausal in the time asymmetric phase, as noted by the low occurrence of zeros, and progressively evolves towards domination by causal moves at late times, in the symmetric phase.
In \textbf{(b)} we show the frequency of zeros, computed in bins of 40 elements, in the evolution of the causal set. This figure confirms the indication of the left panel of a rapid increase in the frequency of zeros towards late times, denoting the decrease in discausal jumps.
\label{cycle_full}}
\end{center}
\end{figure*}

Given that a typical ECS limit cycle is composed of only two elements, this means that during a limit cycle the system returns to the same spatial position with every second move. So if we subtract every second element we should get zero when the system is in a limit cycle, and not zero when not.

In the last panel of Figure~\ref{zoom_in} we see that the jumps outside of limit cycles, are typically jumps highly discausal jumps -- this means that the subtraction of consecutive elements is a proxy for identifying discausality. When the subtraction is non zero the corresponding move is strongly discausal. This is particularly true for the last panel of Figure~\ref{zoom_in}, that is, for late times.

The subtraction is given by $|z^a_I - z^a_{I-2}|$ where $z^a_I$ are the space time coordinates of the $I$-{\it th} event in the causal sequence, and $a$ is the spacetime index. We take $a=1$ to give the 1d spatial dimension.  In Figure~\ref{cycle_full} we plot this difference versus event order $I$.

The goal is to check Figure~\ref{cycle_full} for any evolution of the amount of zero values of the subtraction. As the system evolves from irreversible to time reversible we want to cross-check how does the discausality progress.

Figure~\ref{cycle_full} shows a clear increase of the rate of zeros as the system evolves. At late times the system is strongly in the causal phase, and the amount of zeros is large, meaning the number of discausal moves is at its minimum. There is a direct correlation then, between the degree of time asymmetry in the system and the number of moves that oppose causality.

The fact that disorganized evolution of events in spacetime signals a (weak) violation of causality is perhaps not surprising in retrospect. But this is the first instance where time asymmetric dynamics is associated with moves which apparently violate causality in the embedding of the causal set, where symmetry in time denotes an alignment of the causal set with the time of the embedding.

\subsection{Becoming}

For many years, retrocausality has been a mere interpretation of QM, at times helpful, but never vital.  Here we examine a broader and more fundamental form of retrocausality, and even stronger violations of traditional causality, namely disordered causality. Importantly, the appearance of disordered causality is not a matter of assumption here, but rather an
outcome of a very general model –- ECS. Disordered causality in this framework is vital for the complete understanding of emergent spacetime.

The emergence of spacetime from events follows earlier advances in physics, where what appears to be a fundamental ingredient of physical reality was shown to emerge from a more primitive one. Mach, for example, considered space and time as secondary to masses. We now suggest a dynamic version of this idea.

Some speculative and very general versions of this idea have already been proposed. Elitzur and Dolev \cite{Becoming1} suggested that spacetime emerges from events via Becoming, namely with the flow of time being granted ontological reality. Instead of the conventional block universe account in which all past, present and future events have  the  same  degree of existence, some  physicists, e.g. \cite{Becoming2}-\cite{SURT} argued  that  the passage of time is not illusory but rather a fundamental property of time\footnote{The debate between the block universe conception of spacetime and the primacy of becoming goes back at least to a famous debate between Albert Einstein and Henri Bergson, in Paris in 1922\cite{debate}.}. The energetic causal set model was indeed inspired by similar ideas \cite{ECS1}.

To illuminate the physics of emergence of spacetime,  here is a simple model, of how becoming can enable spacetime emergence from interactions. This may help illuminate the description of disordered causality proposed in the previous sections.

\begin{figure*}[t!]
\begin{center}
\includegraphics[width=0.9 \textwidth]{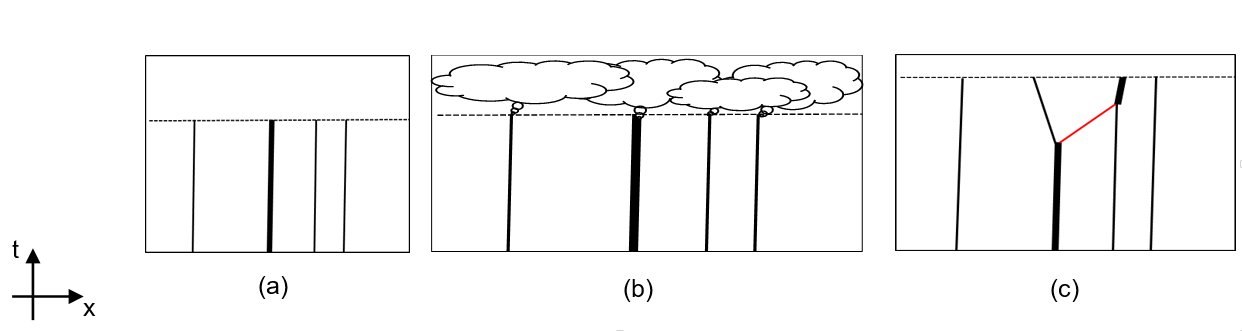}
\end{center}
\caption{The three basic stages of Becoming.
\label{becoming}}
\end{figure*}

Consider a few atoms of which one is excited. Eventually it will emit a photon that will be absorbed by another, ground-state atom.  Classical physics allows only one time-ordering, i.e. one emission, followed by one absorption, plus several non-absorptions which, being interaction free measurements \cite{IFM}, also count as part of the interaction \cite{Counter}. Taking spacetime as emergent, however allows a much richer picture. Here, the past is fixed like in the conventional block universe view, world-lines and all. The future, in contrast, does not yet exist. Spacetime thus ``expands,'' like in conventional cosmology, into the future, but this allegedly happens at each instant of time. in ``pre-spacetime'' which, as can be seen in Figure~\ref{becoming}, is ``beyond'' the edge of the fixed past.

Figure \ref{becoming}(a) shows the world-lines of the excited and ground-state atoms (thick and thin, respectively), and the moving ``now'' drawn as a broken line, made straight for convenience with no regard to relativistic simultaneity. ``Beyond'' the Now's edge, interactions take place in the pre-spacetime, and causality is still disordered. The exchange of energy and momentum between them is made first in this pre-spacetime, represented, in lack of a better symbol, as clouds in Figure \ref{becoming}(b) ``Next'' (in the deeper temporal sense), and following the process previously described in \cite{ECS1}-\cite{ECS3}, a new spacetime region is formed in Figure \ref{becoming}(c) with its events, in accordance with relativistic causality.

The atoms' relative positions and momenta within this new spacetime origin are determined by the pre-spacetime interaction. The consequences seems tempting: Electromagnetic interactions like attraction/repulsion mediated by photons, as well as gravitational attraction mediated by gravitons, even the Lorentz transformations of length and clock time --- all form as natural consequences of the pre-spacetime interactions.
This is of course highly speculative and lacking in detail, presented here only as an illustration to the possibilities opened by dynamizing spacetime. In the work in this article we have taken a much more prudent and simpler model exhibiting retro- and disordered-causality, namely ECS.

\section{Conclusions}

This work completes and extends the work in the companion paper~\cite{CCES1} with regards to types of retrocausality and their role in both foundational quantum mechanics and quantum gravity. We have identified two very different types of possible violations of causality,
\begin{enumerate}
\item Strong, or time symmetric retrocausality, which tries to restore the time symmetry that is broken by the irreversibility of the usual quantum measurement postulate. This was is the retrocausality addressed in~\cite{CCES1} and is further discussed, e.g. in \cite{R3}.

\item Weak or disordered causality or discausality, which describes the phenomena we recently discovered within the ECS model.
\end{enumerate}

More specifically, the ECS model exhibits three notions of time:

\begin{itemize}
\item[] A.  Birth order, which is a total ordering, representing the full non-local causal dynamics. This represents the fundamental causality underlying energetic causal sets, that motivated the model proposal, and is never violated. 

\item[] B. Causal time: derived from causal relations in the set built upon event by event, relating children to their parents.  This is the causal structure obeyed by flows of energy and momentum.

\item[] C. Spacetime time (or emergent time): resulting from embedding the causal process in the emergent Minkowski spacetime.
\end{itemize}

We have seen that Time A  can be in opposition with Time C.  This means that causal time always moves forward in the causality sequence but it can be misaligned with Min\-kowski time. In this picture the fundamental causality (birth order) is still fundamental, irreversible and asymmetric, but it can sometimes run with, and sometimes against the direction of time in the macroscopic and emergent spacetime.

In previous works we have started from the premise that irreversibility of time is fundamental. In this work we found out that the algorithm for generating new events in ECS allows for weak apparent violation of causality in the emergent macroscopic spacetime.

Therefore, these new results lead us to a new formulation of that premise: irreversibility of time in the form of a causal arrow in the set (Time A) is fundamental, but in the emergent spacetime there can occur sequences of events which apparently violate the direction of time (Time C) as measured by the emergent spacetime geometry. Another formulation states that the time ordering of events in manifold time may be reversed, but the underlying causal order is always uni-directional, and never violated.

We note that the possibility of separating the three notions of  causal order relies on the theory being background independent, so that classical Minkowski spacetime is emergent and dynamically generated.

With this new insight into the properties of causality we can now summarize the properties of fundamental time in the framework of ECS:
\begin{enumerate}

\item Time, as in causation, (Time A), always moves forward in the direction of causal propagation. This, however can be in opposition to the direction of time in the emergent Minkowski spacetime.

\item The edge of spacetime does not move forward in slices of simultaneity: the causal corridor of the set can zig-zag back and forth in spacetime.

\item Misalignment in the structure of causality and spacetime, allows for causal connections to go forward and backward in spacetime. Causally, it goes always forward,  but it can go backward in the emergent spacetime, i.e., \textit{there is just one direction of causation}.


\end{enumerate}

There is  an important point to be addressed when discussing violations of causality or of time's direction, which is whether closed causal loops (forbidden by general relativity) can occur, because it becomes possible to
send signals from the causal future 
into the causal past.
The answer is no, closed causal loops are forbidden in the fundamental global causal order, Time A. The type of causality violations we discuss in this paper does not allow for causal loops or any signalling going backwards in the causality sequence of total order, Time A.

The order of causality in the causal network is never violated. By construction, a causal link, once laid down, has a well-defined direction that is never reversed. Causal links may not be traversed in the opposite direction. Also, an event gives birth to only a finite set of descendants.  Once these are exhausted that event is no longer available for interaction.   It is buried in the past structure of the causal set and will not again come back to the set of events that form the present and generate new events. So a closed time loop is excluded.

A simpler argument contributing to the same proof, is the fact that in ECS one of the fundamental rules is that all events are unique and are never repeated. For the particular model we chose to simulate in \cite{ECS1} for reasons of computational performance, this is ensured by assigning to each event a real number, which is determined by its progenitors and represents its energy-momentum. Since part of the determination is probabilistic, and has as an outcome a real number, an event can never be repeated or occur more than once. Together these arguments forbid the existence of closed time like curves.

We also hypothesized that retrocausality of type 1 could be the origin of quantum non-locality, giving rise to processes which when seen in {\it space} are non-local and non-connected, but revealed to be connected when viewed in the {\it spacetime} picture (see also \cite{R7}). This leads the
realization that the quantum world is more interconnected than one may think.
In future work we will also address the issue of total versus partial order, which we have encountered here in the ECS results.

\section*{Acknowledgements}

We are grateful to Yakir Aharonov and Andrew Liddle for many helpful discussions.

This research was supported in part by Perimeter Institute for Theoretical Physics. Research at Perimeter Institute is supported by the Government of Canada through Industry Canada and by the Province of Ontario through the Ministry of Research and Innovation. This research was also partly supported by grants from NSERC and FQXi. M.C.\ was supported by Funda\c{c}\~{a}o para a Ci\^{e}ncia e a Tecnologia (FCT) through grant SFRH/BPD/111010/2015 (Portugal). LS and MC are especially thankful to the John Templeton Foundation for their generous support of this project. Further, this work was also supported by Funda\c{c}\~{a}o para a Ci\^{e}ncia e a Tecnologia (FCT) through the research grant UID/FIS/04434/2013.

\section*{Appendix - Robustness of simulations}

In this section we present simulations of the same model in order to test robustness of the results to variation of initial conditions, as well as the response to stochastic evolution. We show 3 runs of the same model discussed in the text. All runs have the same model parameters, i.e. the number of families is 20, and total number of events is $10^4$. The variation in each simulation is then arising from
\begin{enumerate}
\item The choice of initial conditions: 20 randomly selected initial spatial positions, one for each of the 20 families,
\item  The variation in the stochastic component of the dynamics along the evolution of the causal set.
\end{enumerate}

\begin{figure*} [h!]
\begin{center}
$\begin{array}{cc}
\includegraphics[width=0.48 \textwidth]{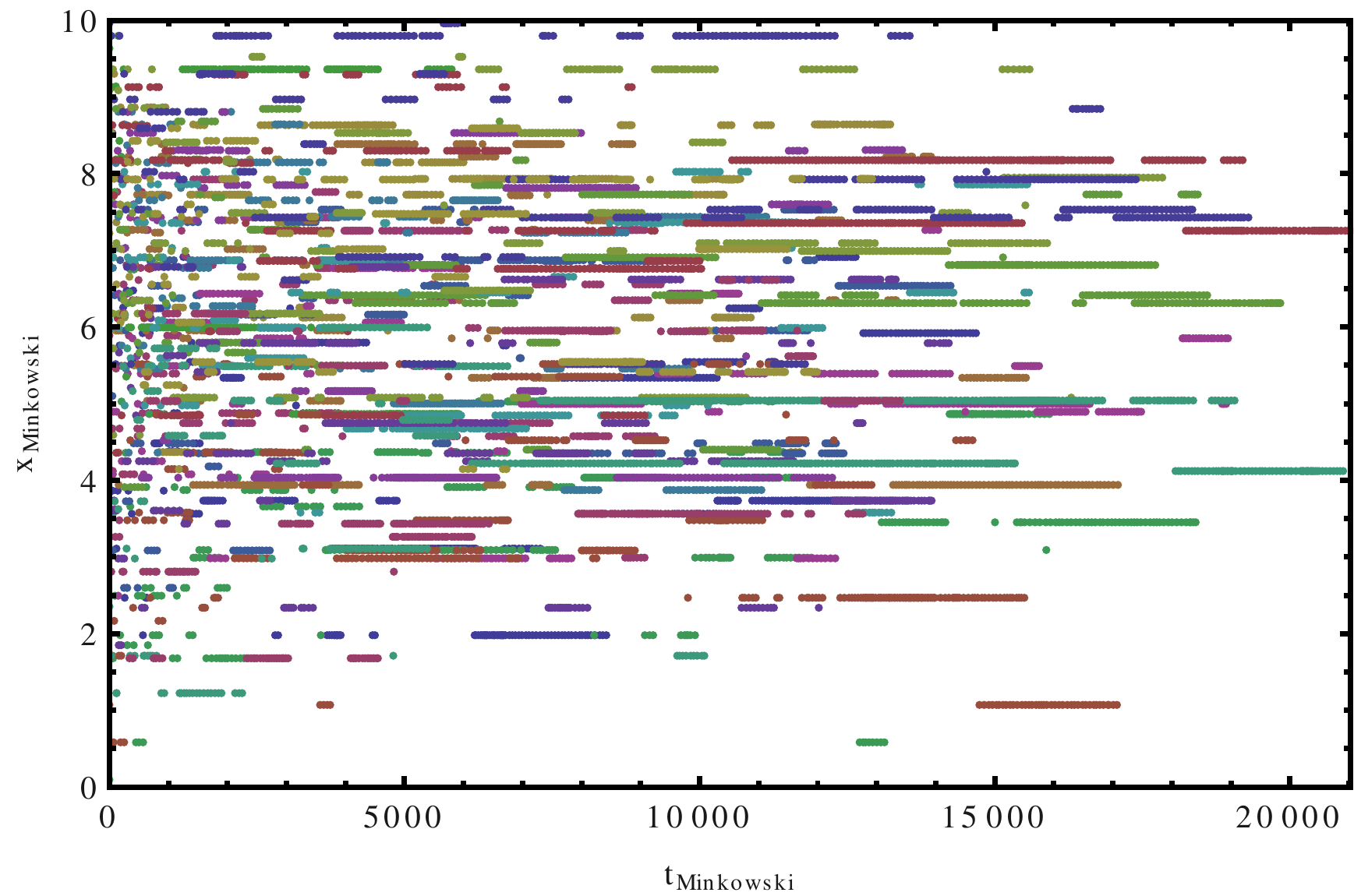}&
\includegraphics[width=0.48 \textwidth]{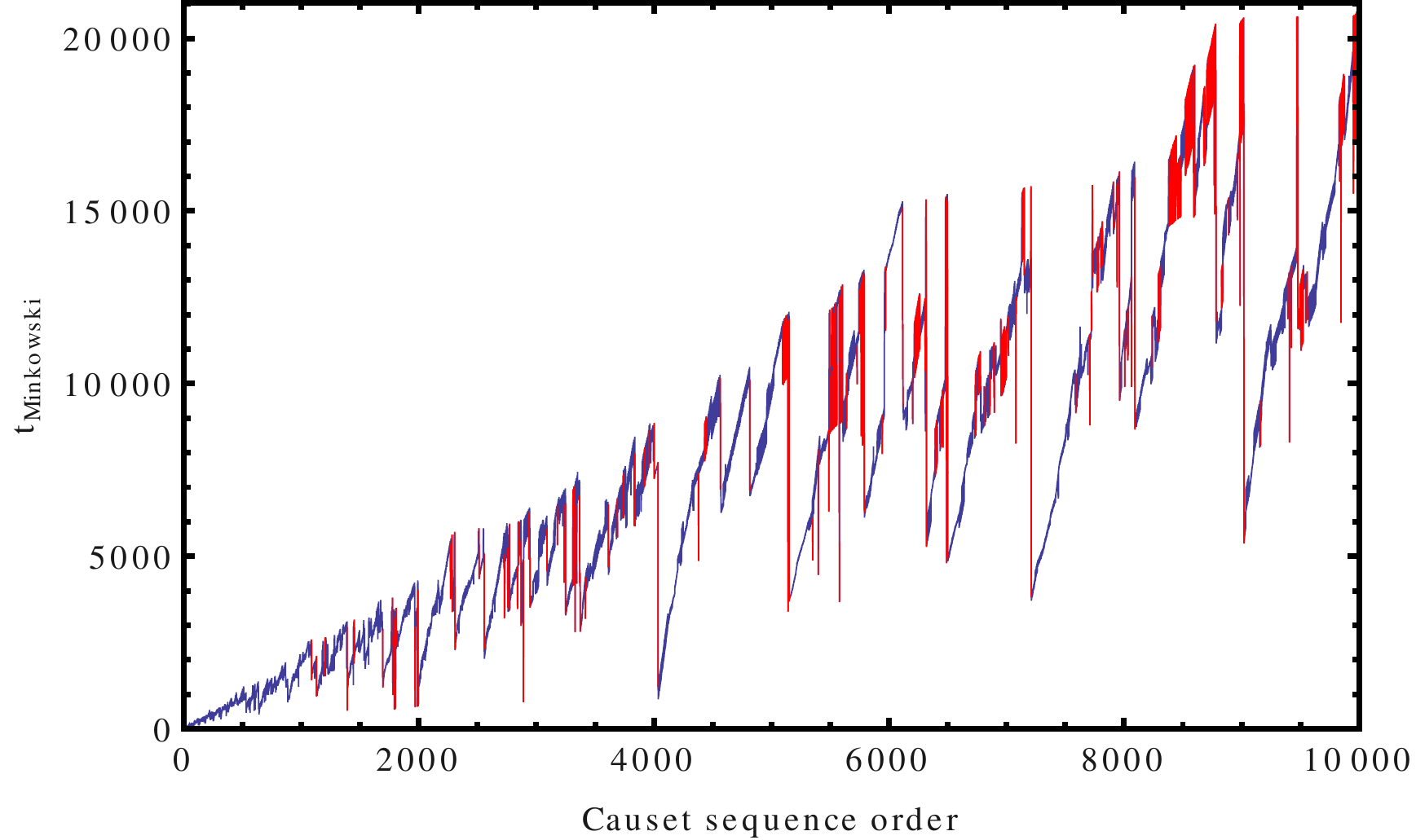}\\
(a) & (b)\\\\
\includegraphics[width=0.48 \textwidth]{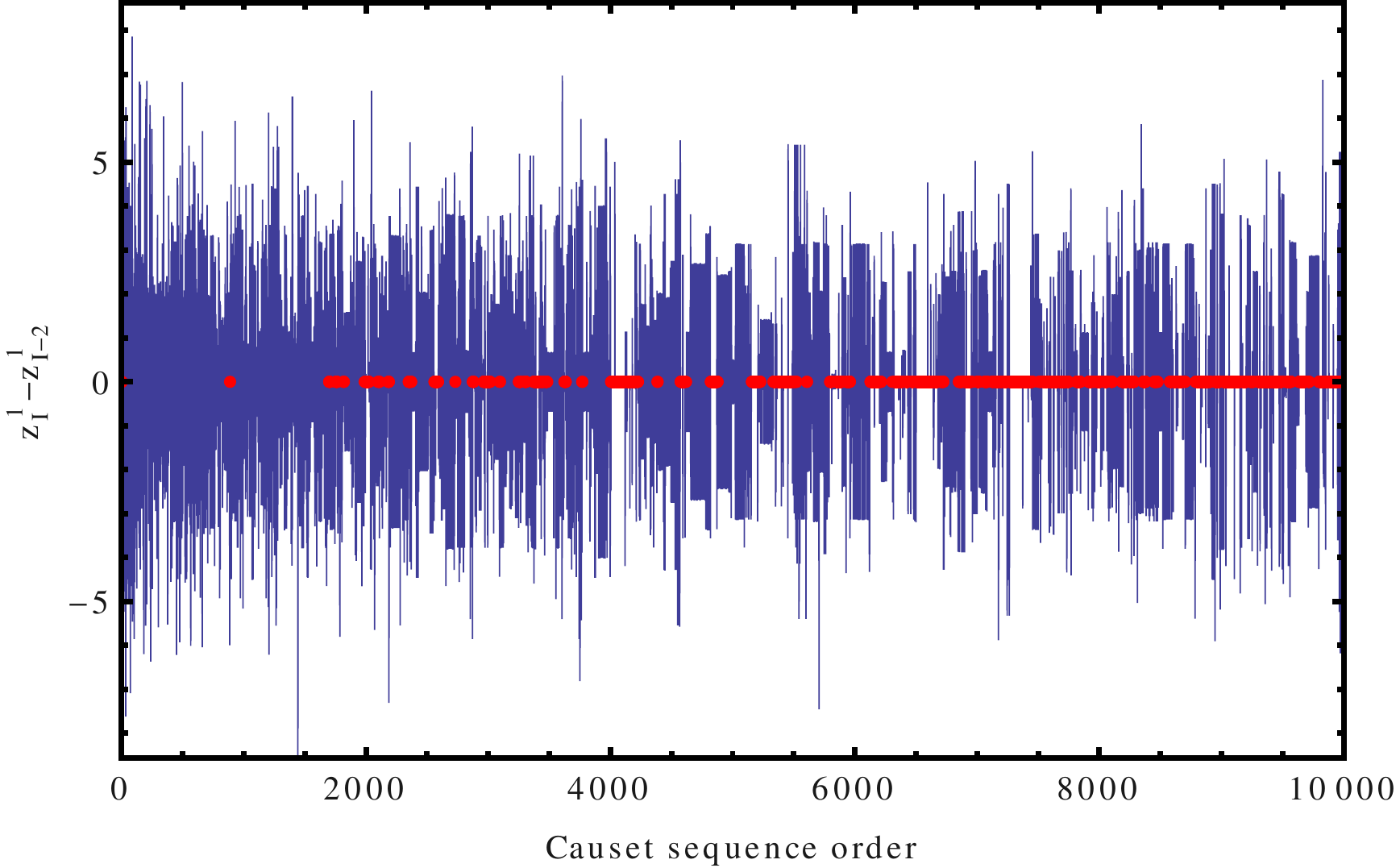}&
\includegraphics[width=0.48 \textwidth]{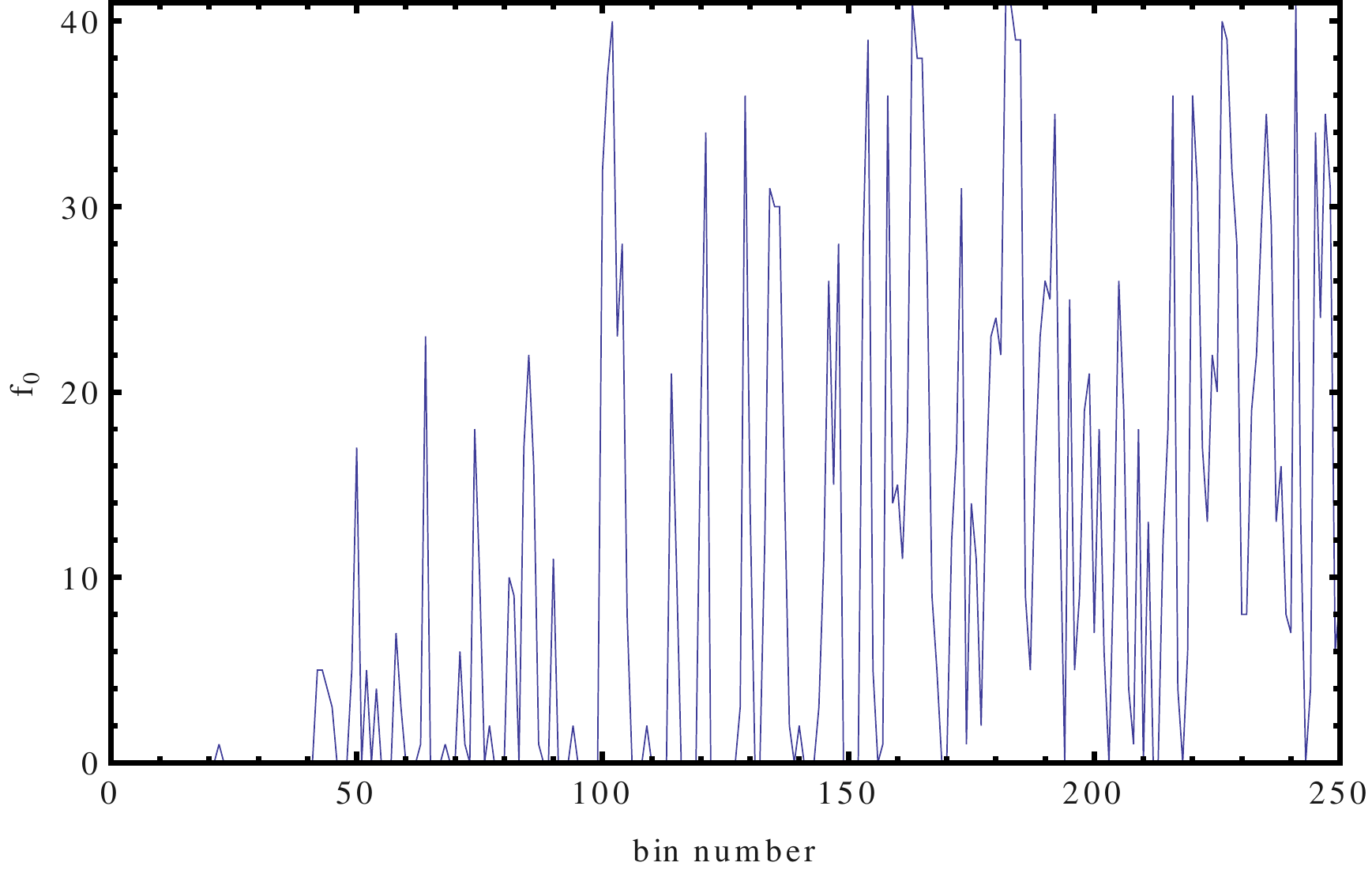}\\
(c) & (d)
\end{array}$
\caption{Simulation I: Robustness of results to variation of initial conditions. Simulation with the same model parameters and different 20 initial spatial positions of each family. Panels show {\bf(a)} Events in emergent spacetime. {\bf(b)} Evolution of Minkowski time with event number. {\bf(c)} Difference between two consecutive elements for detecting the presence of limit cycles - when the difference is zero. {\bf(d)} Evolution of the amount of limit cycles given by the amount of zeros of the previous plot.
\label{app11}}
\end{center}
\end{figure*}
\begin{figure*} [h!]
\begin{center}
$\begin{array}{cc}
\includegraphics[width=0.48 \textwidth]{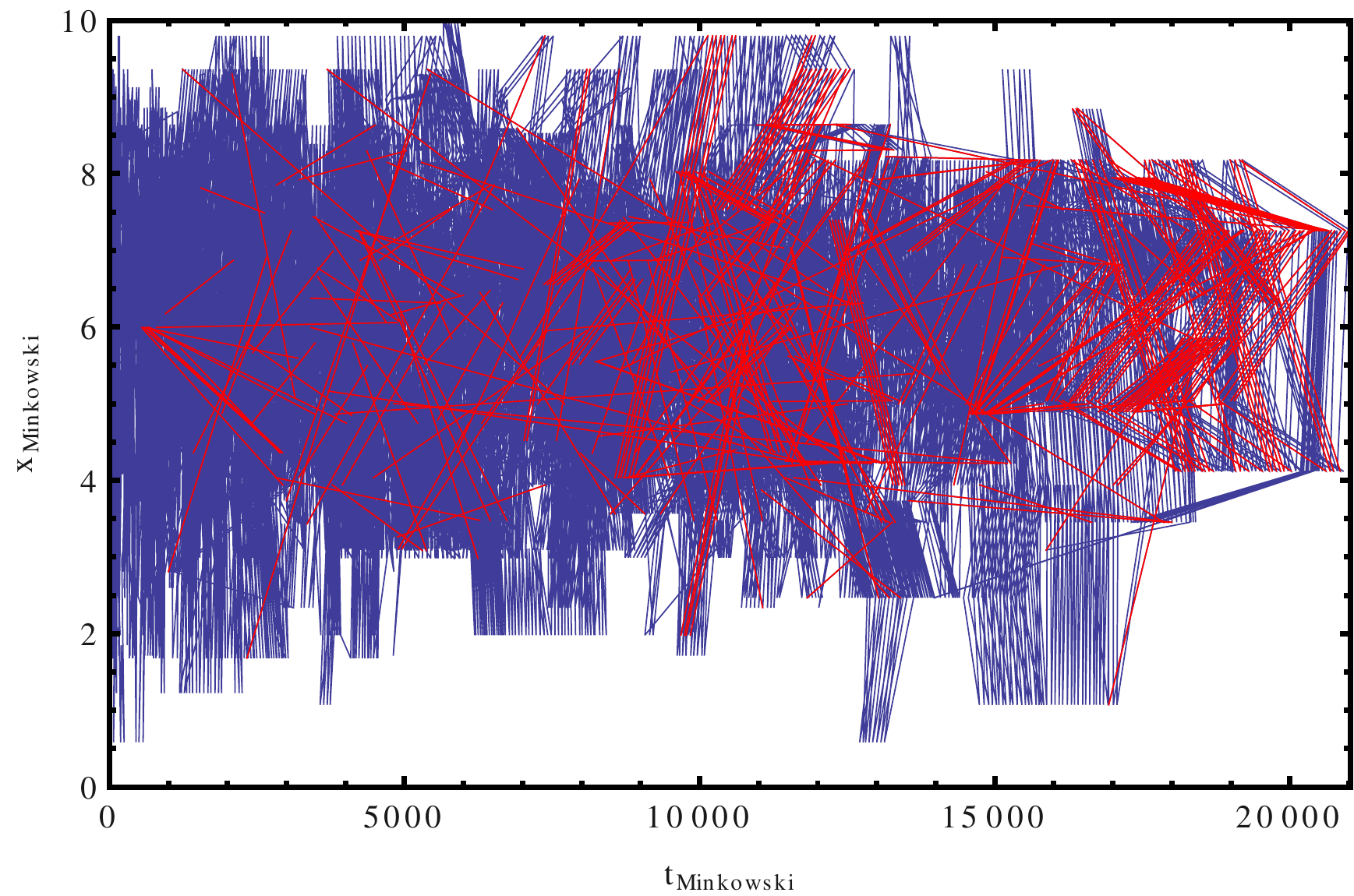}&
\includegraphics[width=0.48 \textwidth]{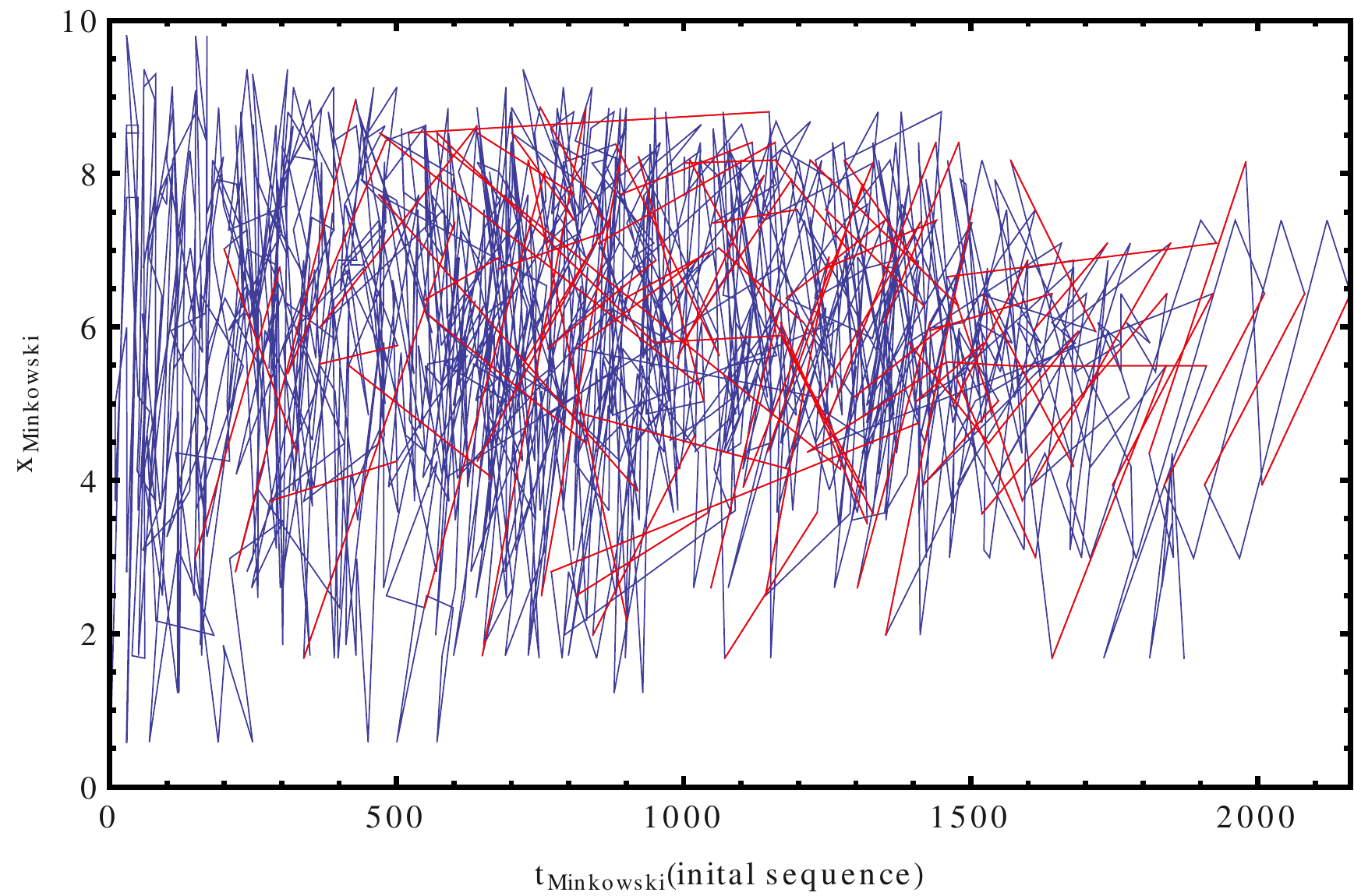}\\
(a) & (b)\\\\
\includegraphics[width=0.48 \textwidth]{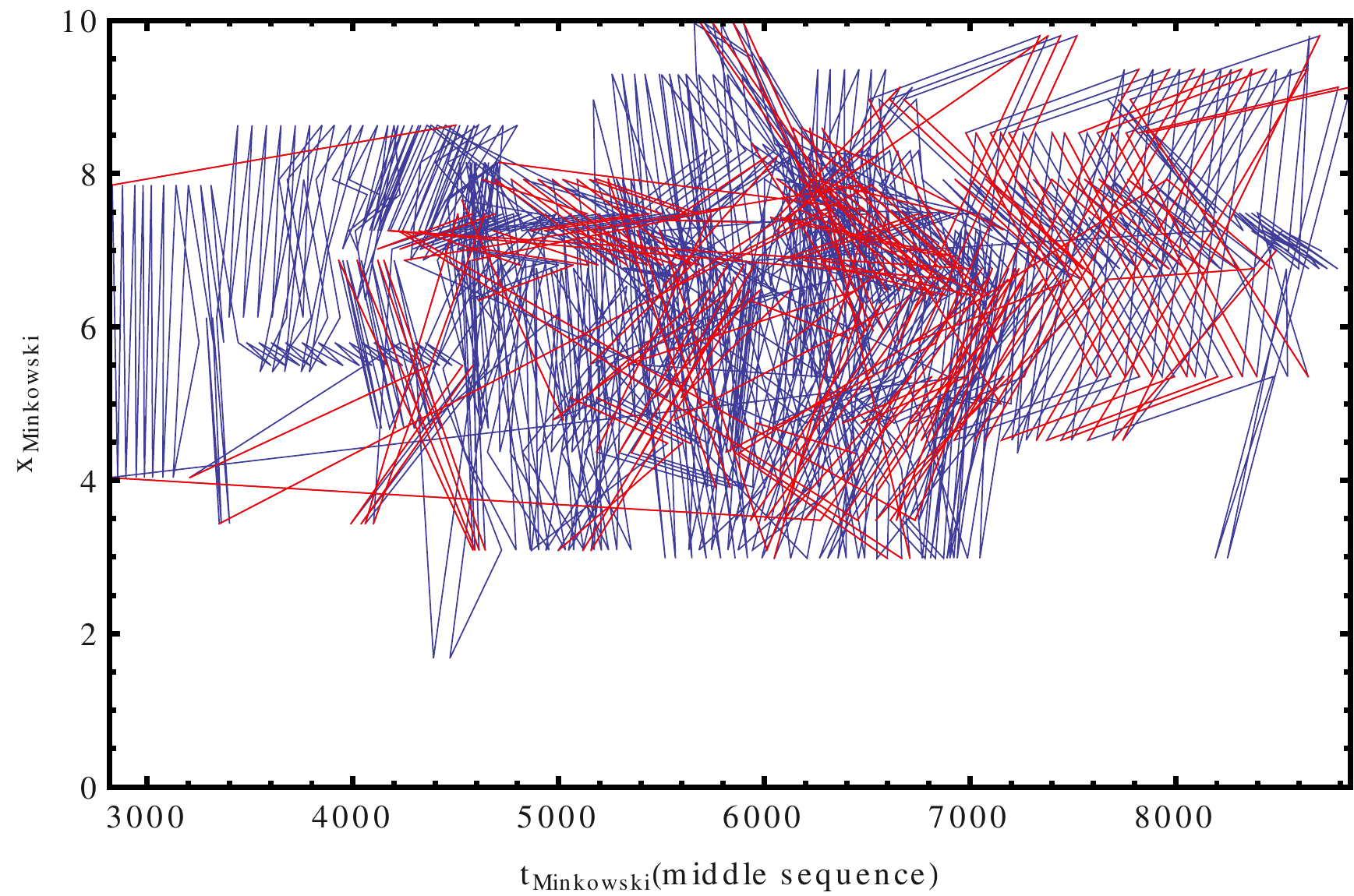}&
\includegraphics[width=0.48 \textwidth]{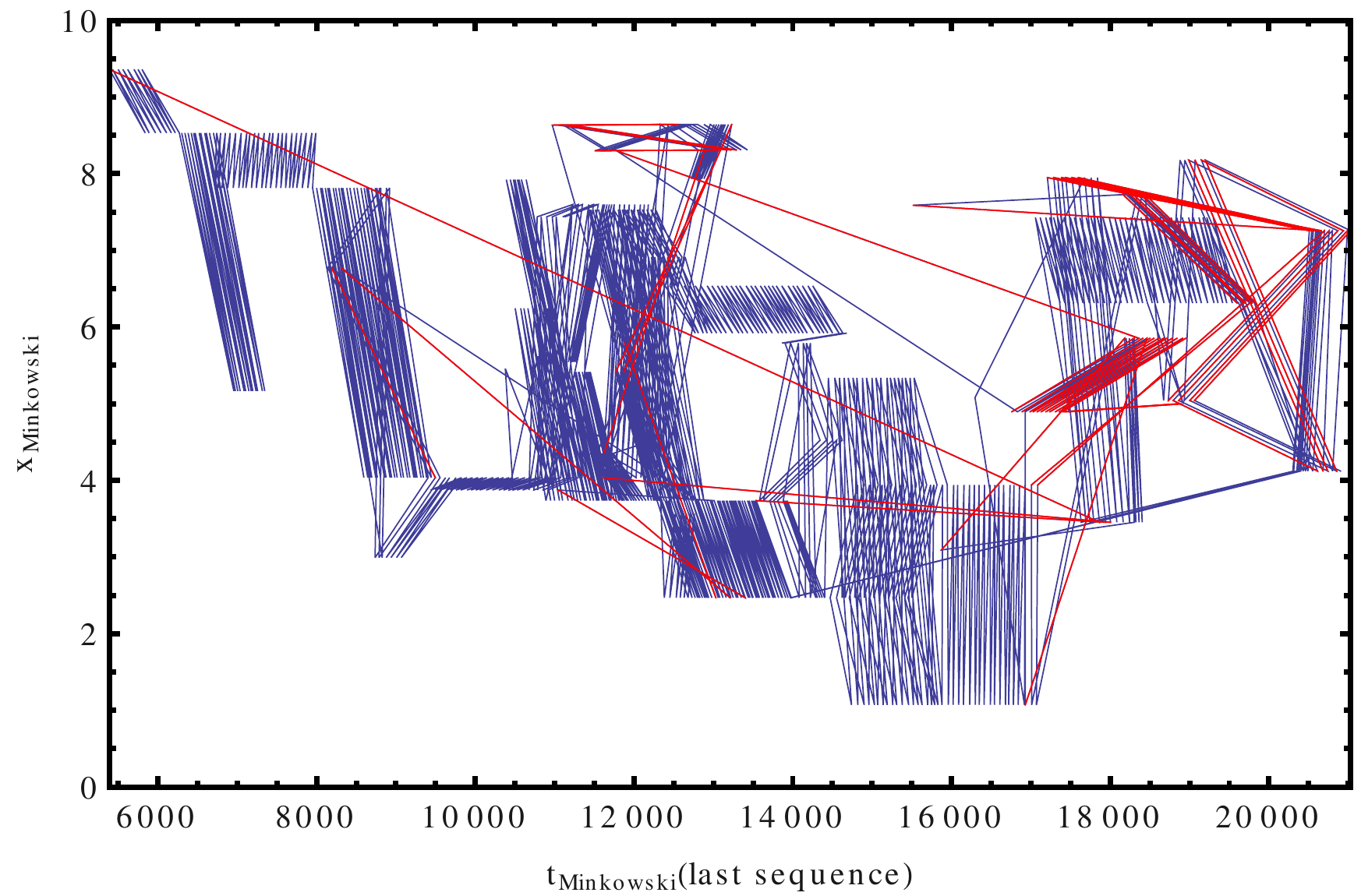}\\
(c) & (d)
\end{array}$
\caption{Simulation I: Robustness of results to variation of initial conditions. Simulation with the same model parameters and different 20 initial spatial positions of each family. Panels show the network of events in the emergent spacetime connected in order of sequence in the causal set. {\bf(a)} Full evolution with $10^4$ events. {\bf(b)} Zoom-in of first $10^3$ events. {\bf(c)} Zoom-in of $10^3$ events from the middle of sequence. {\bf(d)} Last $10^3$ events.
\label{app12}}
\end{center}
\end{figure*}

\begin{figure*} [h!]
\begin{center}
$\begin{array}{cc}
\includegraphics[width=0.48 \textwidth]{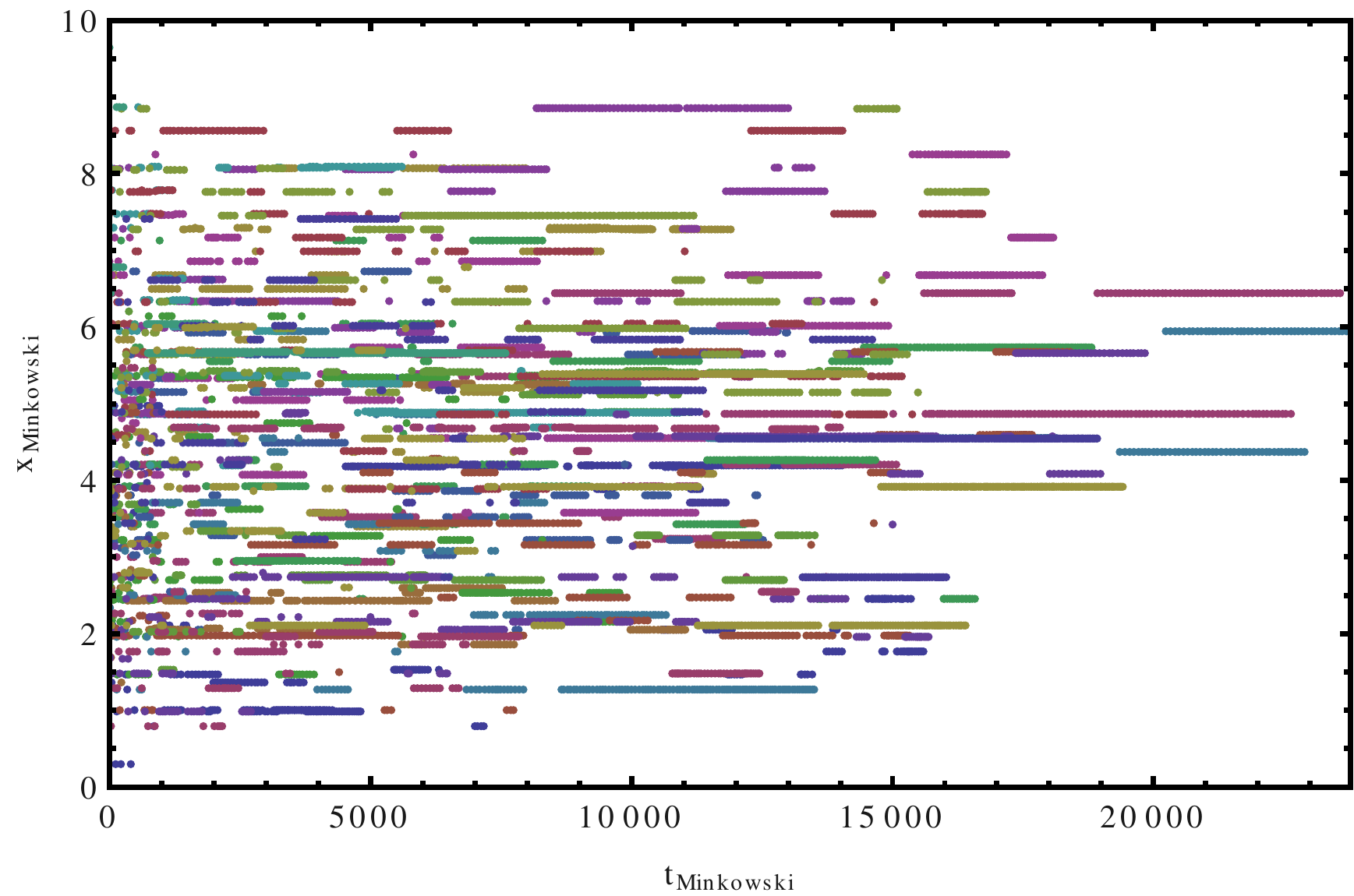}&
\includegraphics[width=0.48 \textwidth]{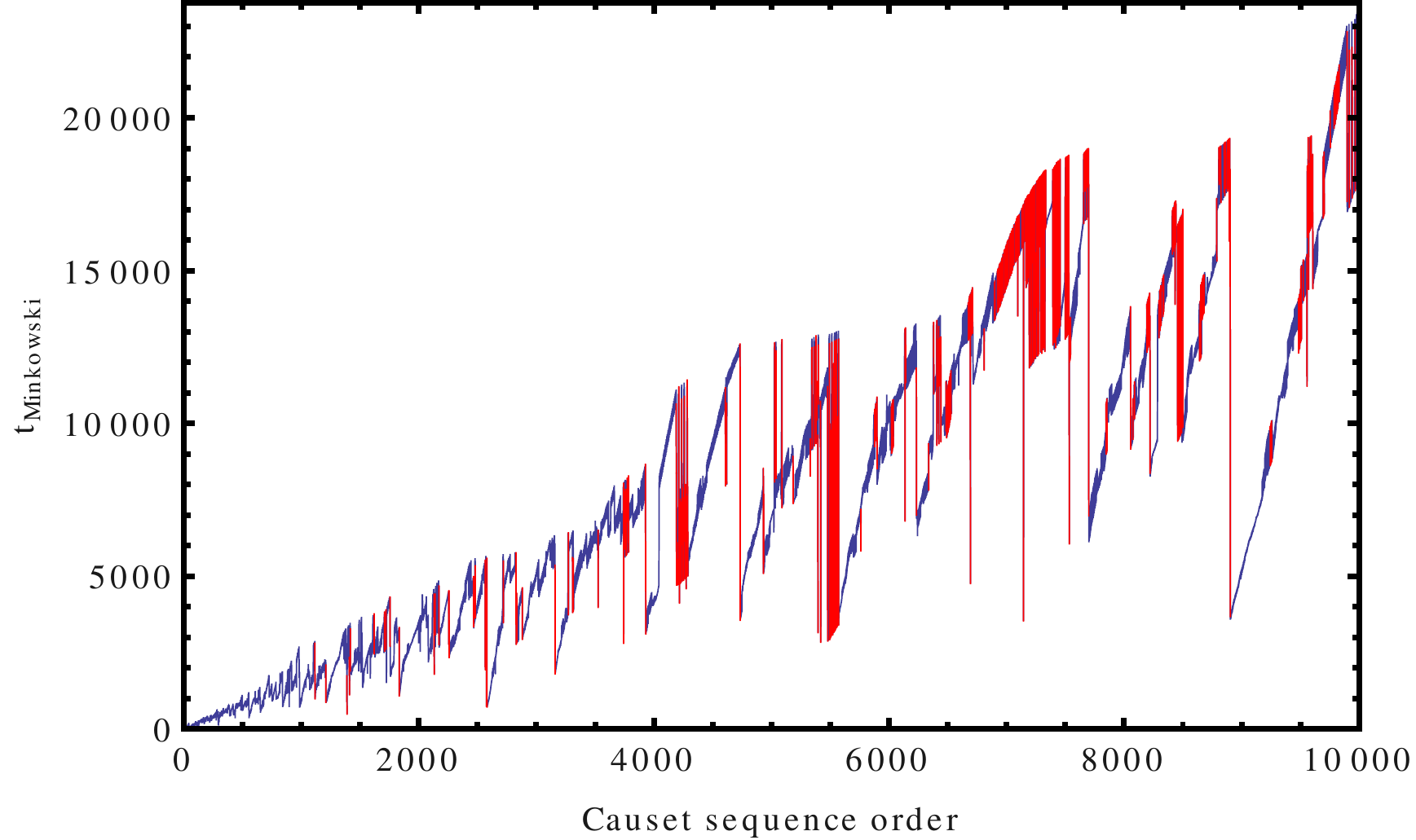}\\
(a) & (b)\\\\
\includegraphics[width=0.48 \textwidth]{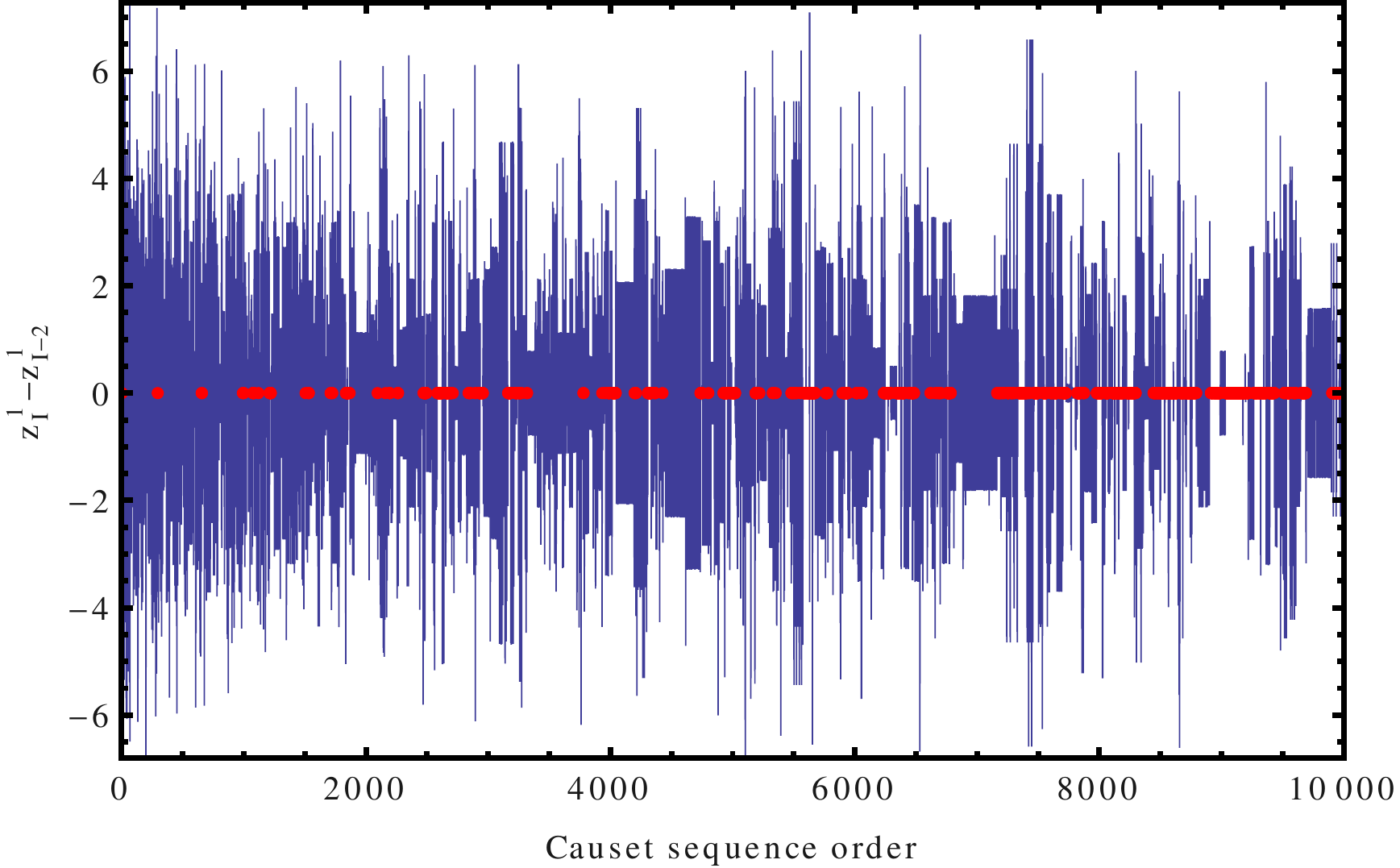}&
\includegraphics[width=0.48 \textwidth]{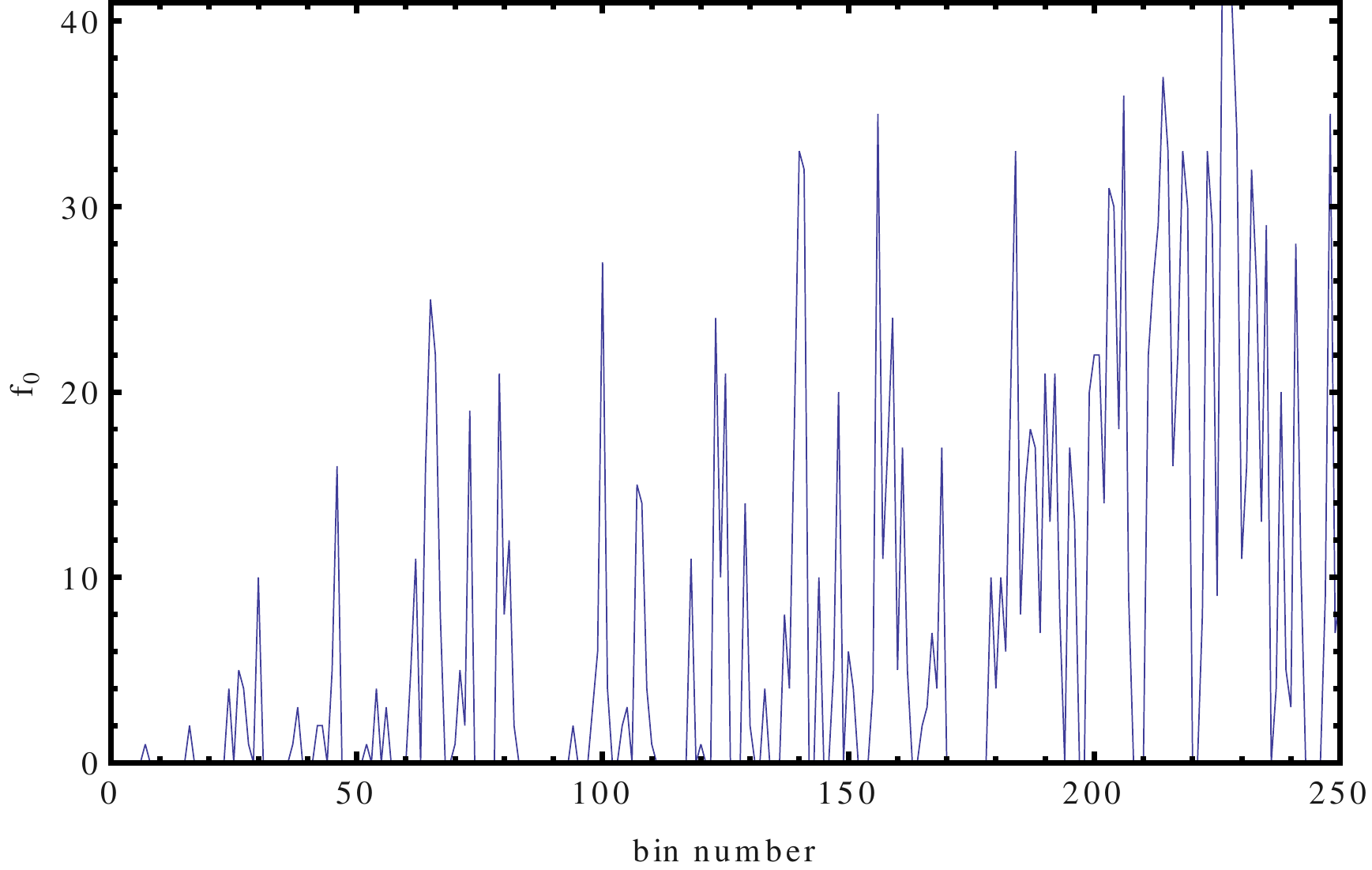}\\
(c) & (d)
\end{array}$
\caption{As Figure~\ref{app11}, for Simulation II.
\label{app21}}
\end{center}
\end{figure*}
\begin{figure*} [h!]
\begin{center}
$\begin{array}{cc}
\includegraphics[width=0.48 \textwidth]{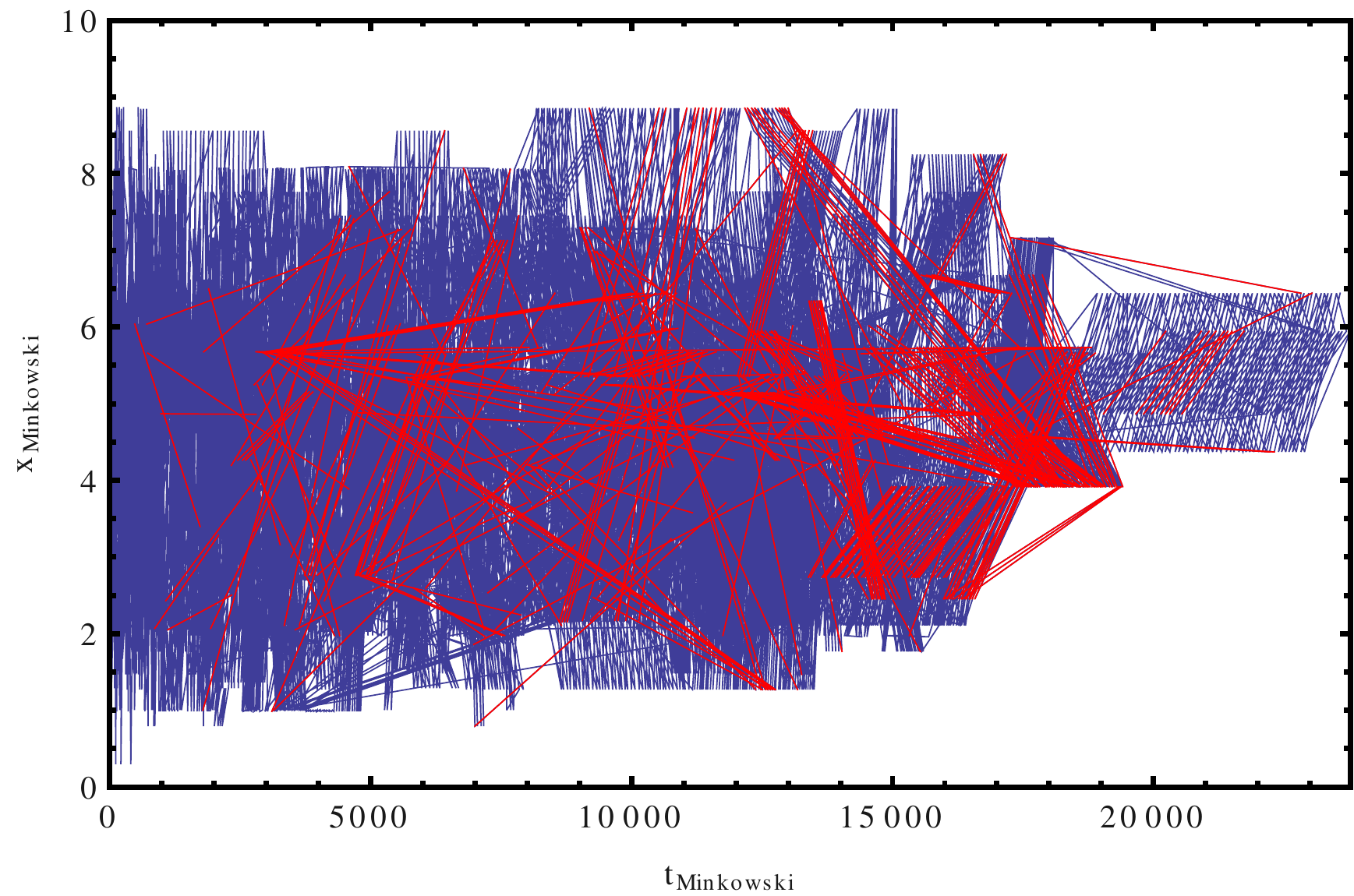}&
\includegraphics[width=0.48 \textwidth]{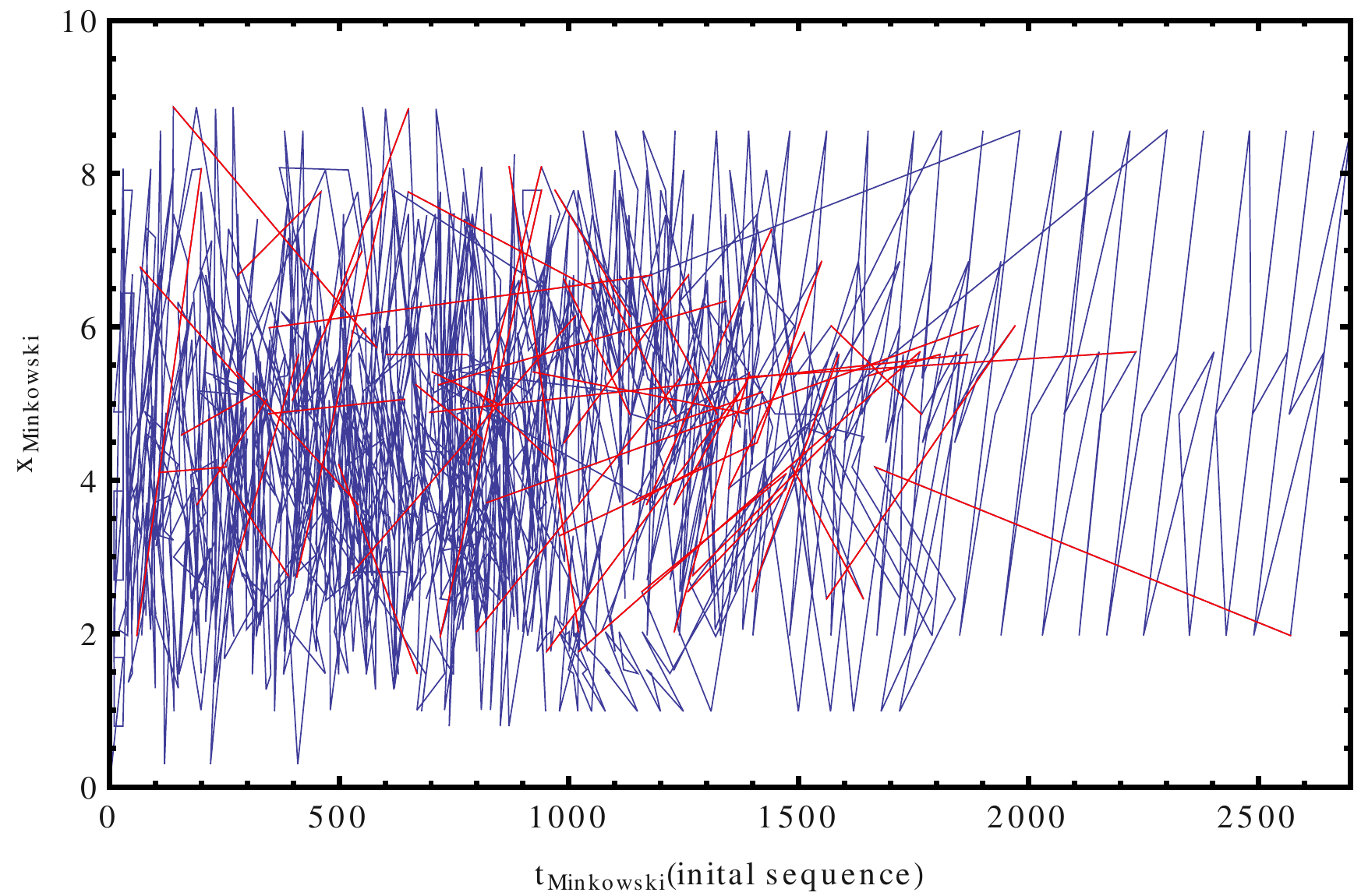}\\
(a) & (b)\\\\
\includegraphics[width=0.48 \textwidth]{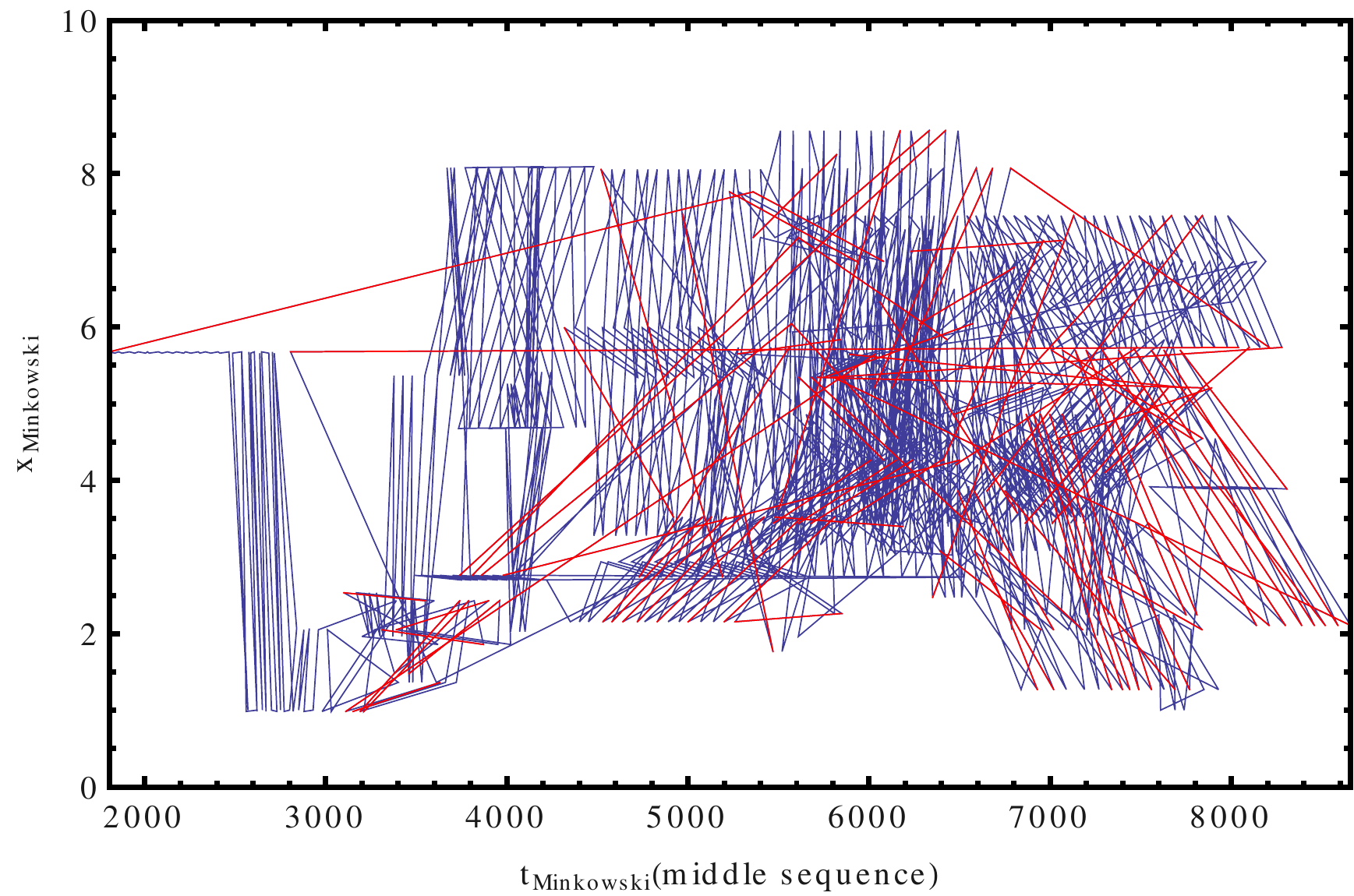}&
\includegraphics[width=0.48 \textwidth]{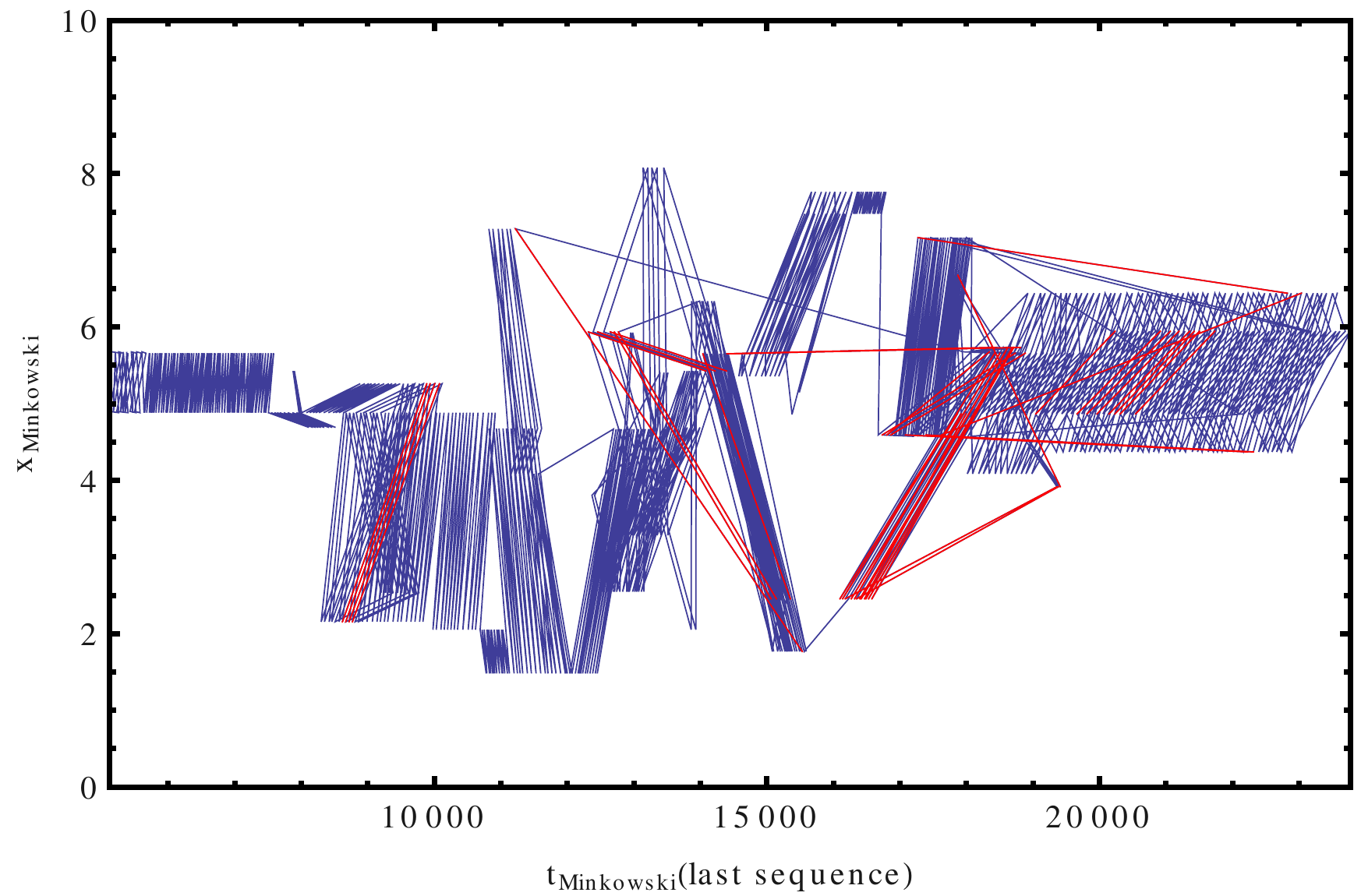}\\
(c) & (d)
\end{array}$
\caption{As Figure~\ref{app12}, for Simulation II.
\label{app22}}
\end{center}
\end{figure*}

\begin{figure*} [h!]
\begin{center}
$\begin{array}{cc}
\includegraphics[width=0.48 \textwidth]{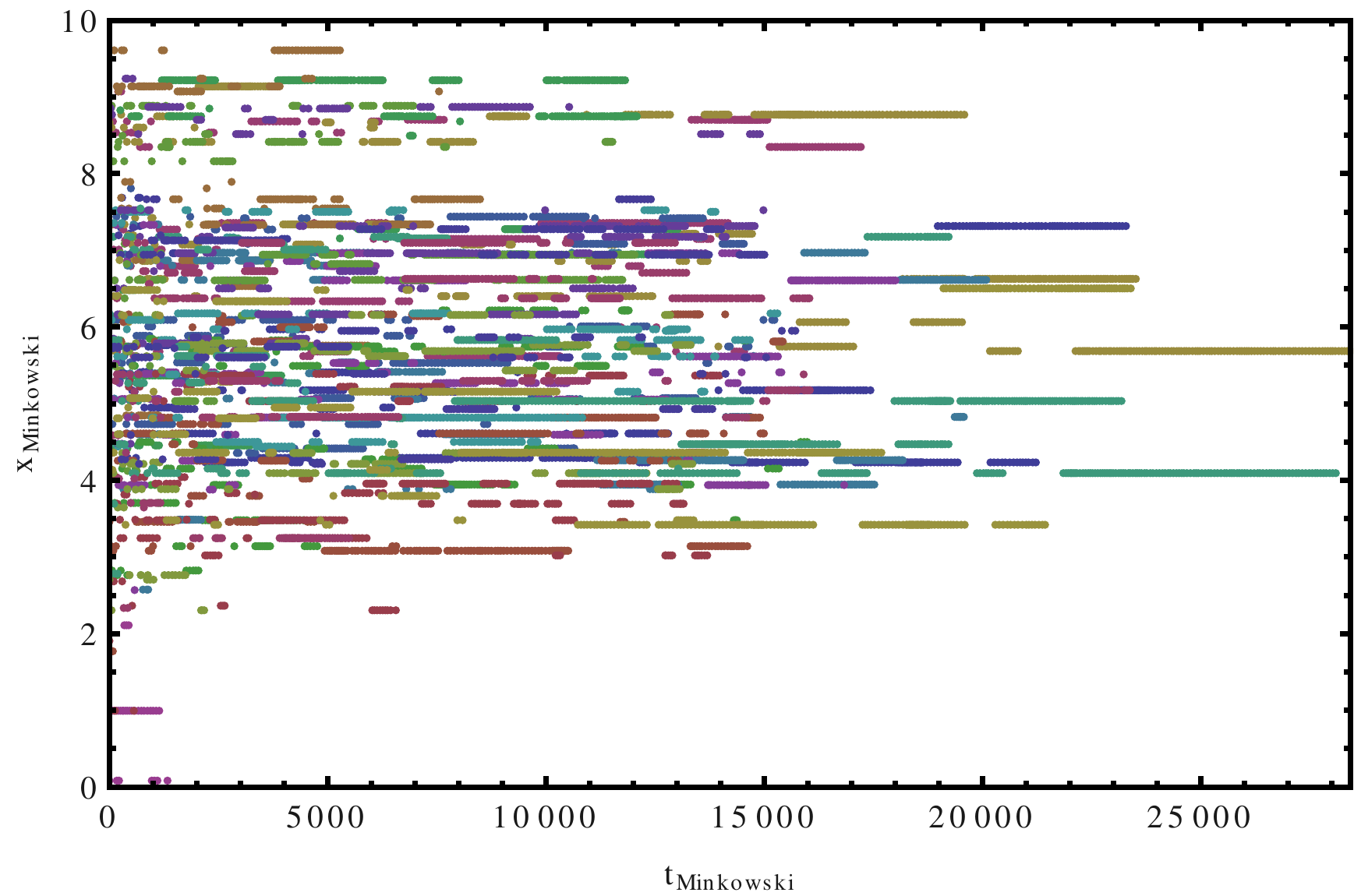}&
\includegraphics[width=0.48 \textwidth]{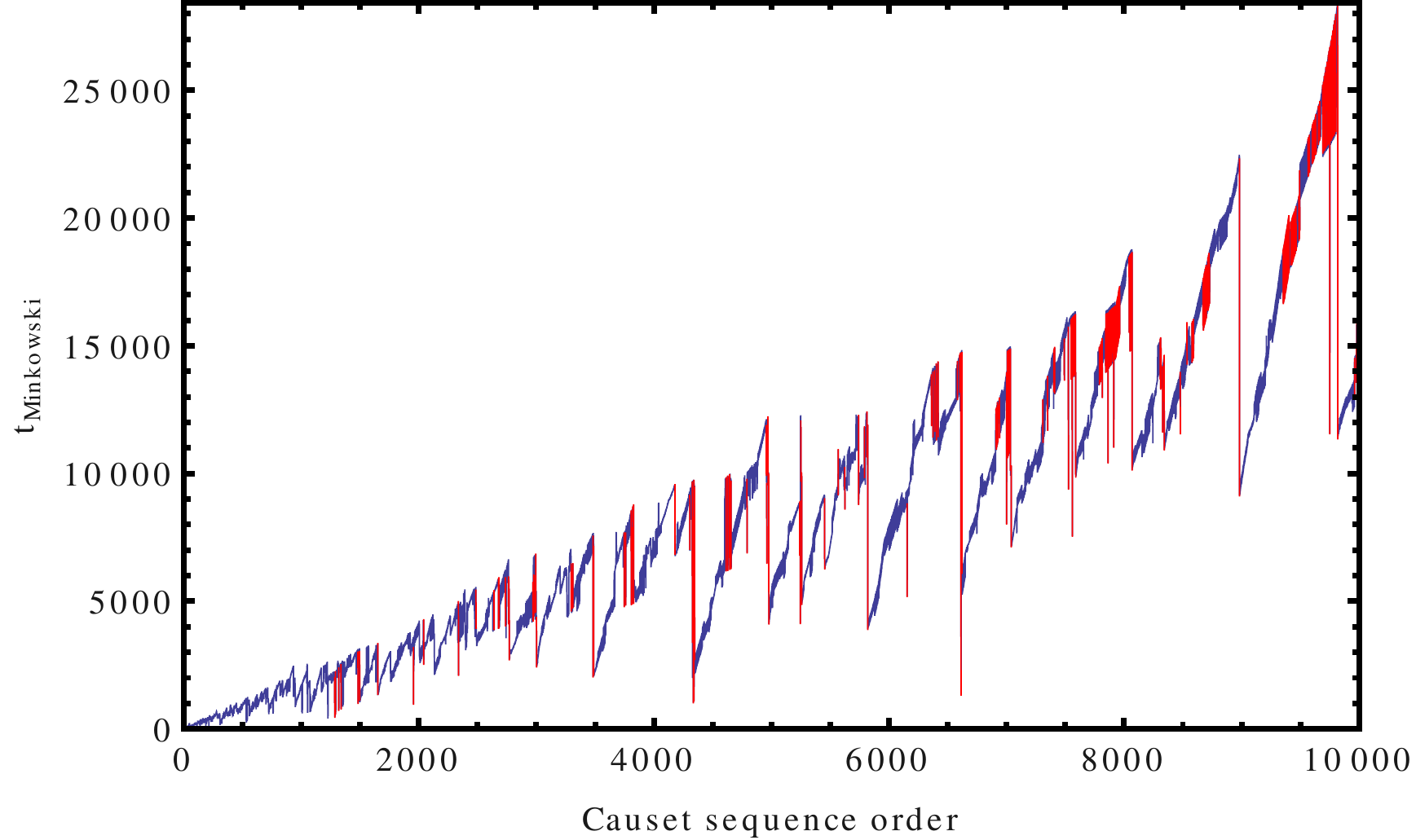}\\
(a) & (b)\\\\
\includegraphics[width=0.48 \textwidth]{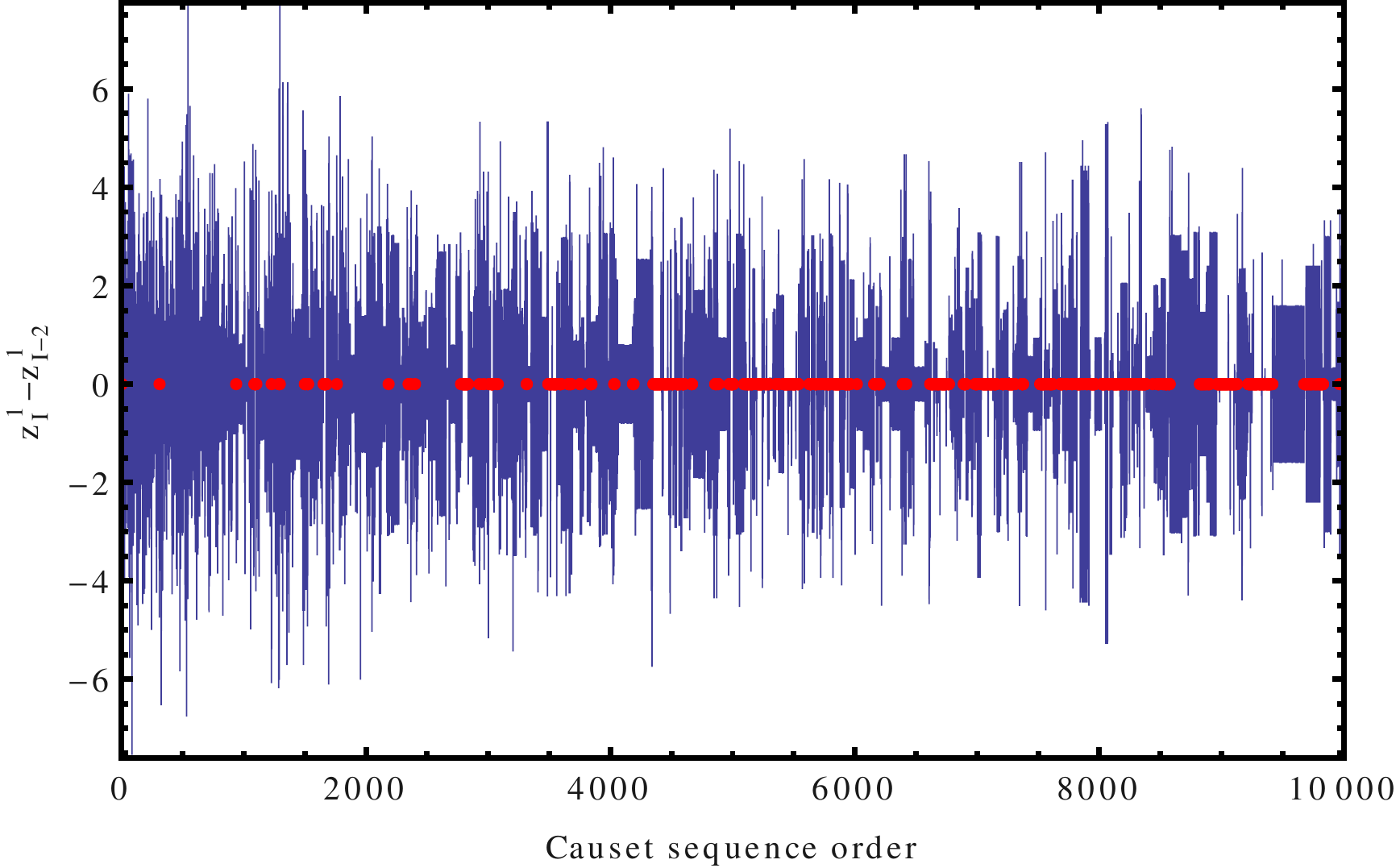}&
\includegraphics[width=0.48 \textwidth]{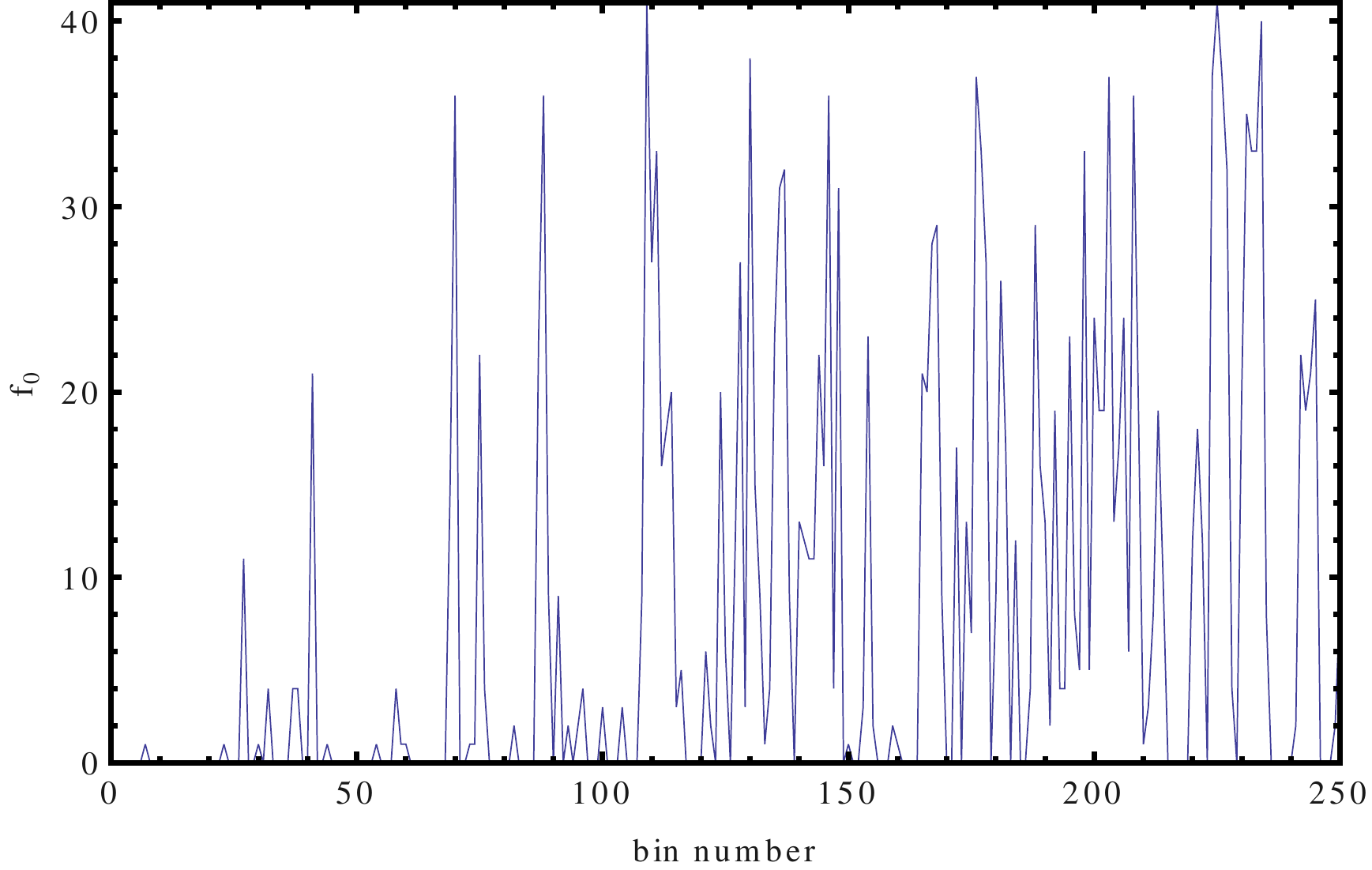}\\
(a) & (b)
\end{array}$
\caption{As Figure~\ref{app11}, for Simulation III.
\label{app31}}
\end{center}
\end{figure*}
\begin{figure*} [h!]
\begin{center}
$\begin{array}{cc}
\includegraphics[width=0.48 \textwidth]{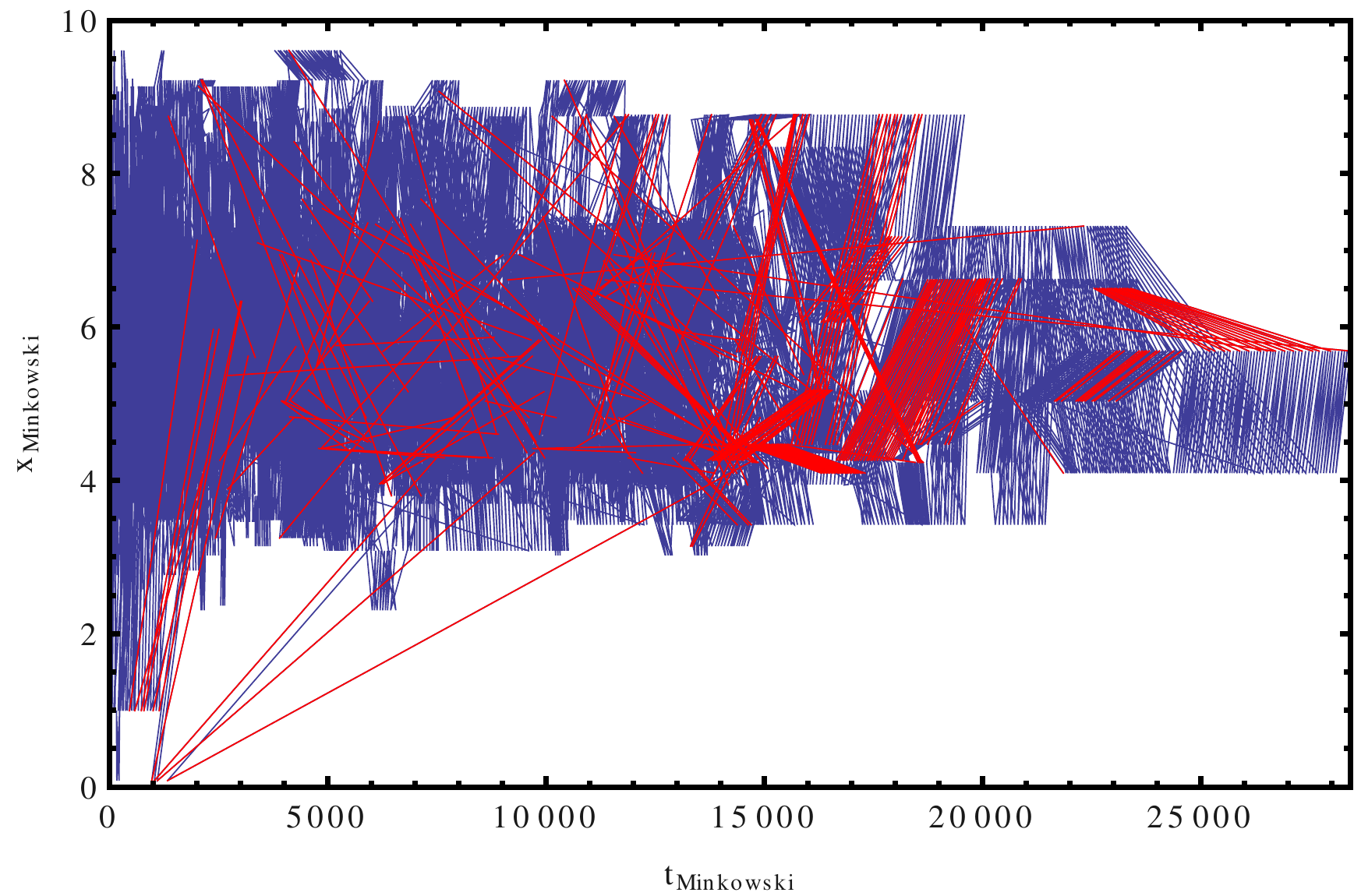}&
\includegraphics[width=0.48 \textwidth]{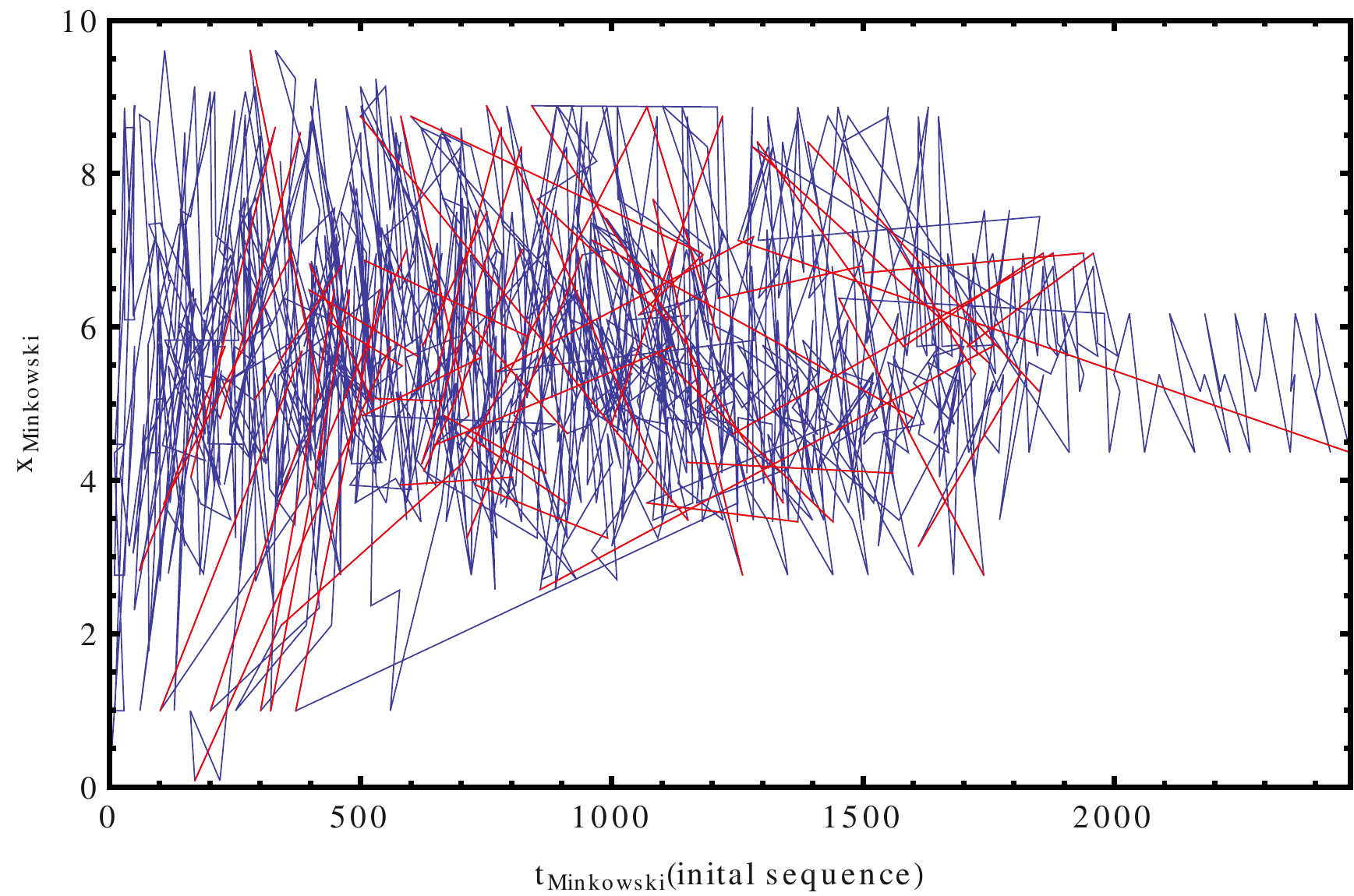}\\
(a) & (b)\\\\
\includegraphics[width=0.48 \textwidth]{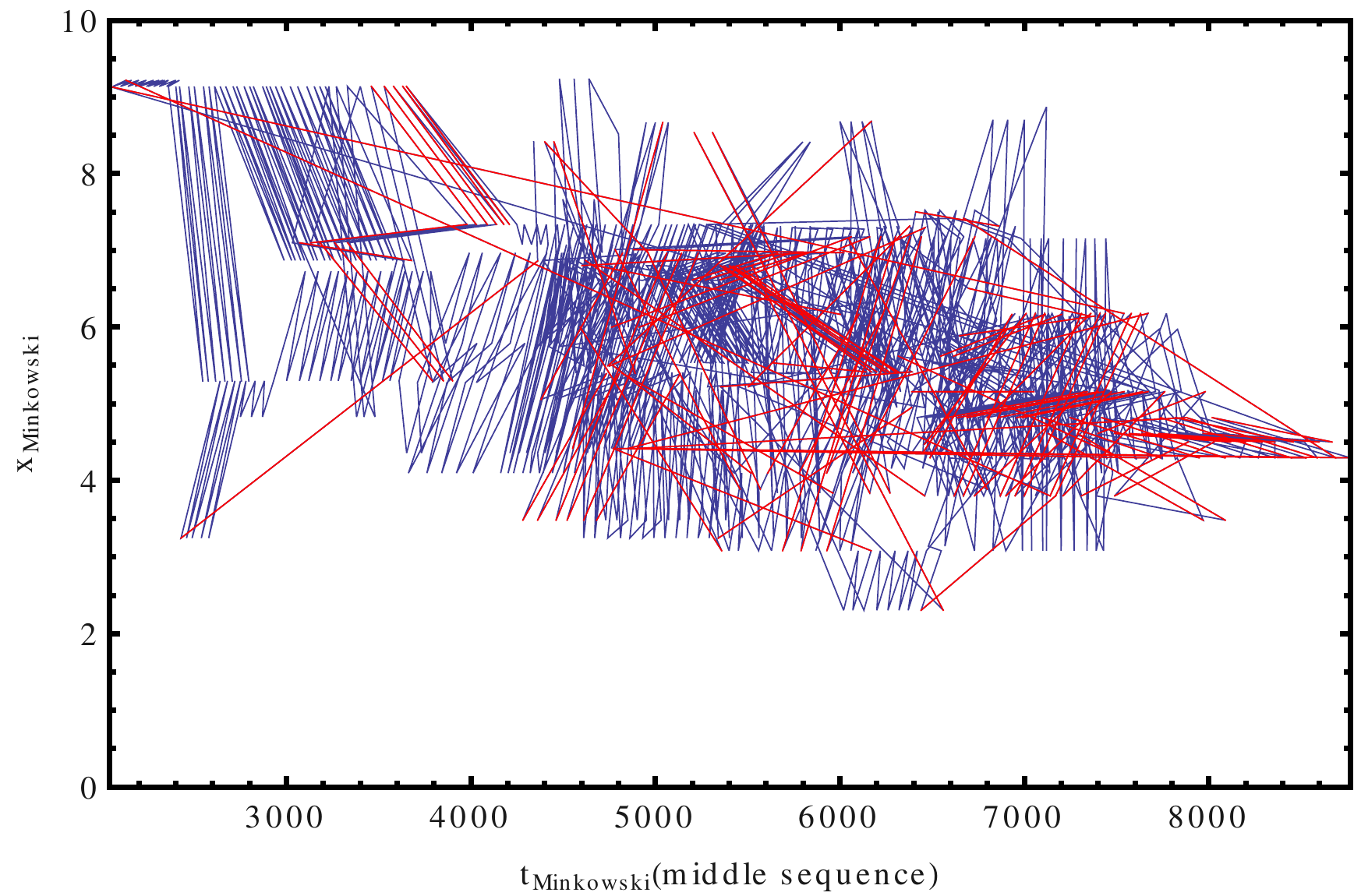}&
\includegraphics[width=0.48 \textwidth]{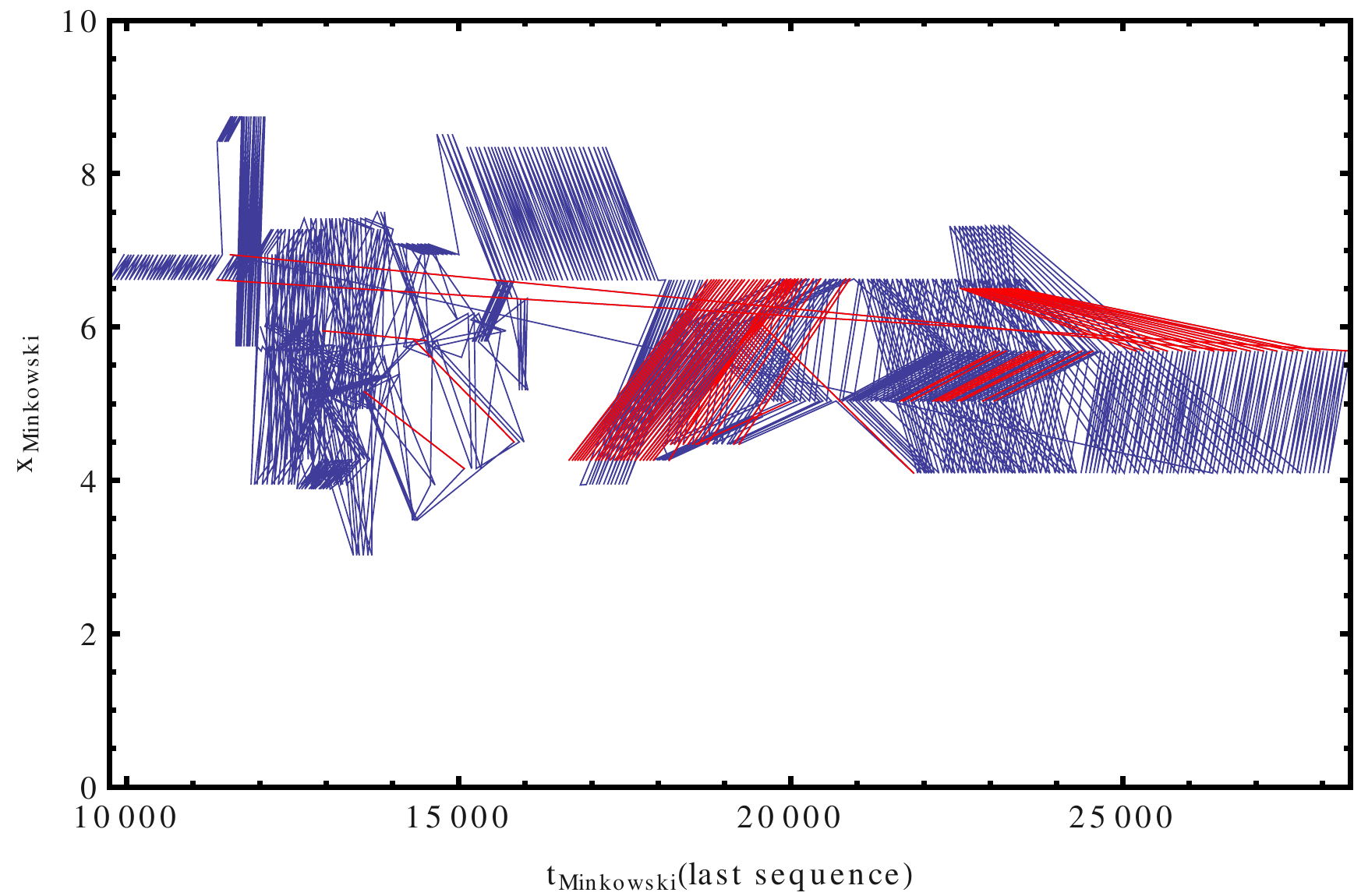}\\
(c) & (d)
\end{array}$
\caption{As Figure~\ref{app12}, for Simulation III.
\label{app32}}
\end{center}
\end{figure*}

We present each run in two sets of figures, one with the emergent spacetime evolution, along with Minkowski time with the causal set evolution and variation with limit cycles. The second set depicts the evolution of the emergent spacetime position versus the event sequence in the underlying causal set: the full evolution is presented along with zoom-in from the beginning, middle, and end of the causal sequence. The discausal jumps are represented in red, as in the main body of the text. Figures~\ref{app11} and~\ref{app12} represent the first simulation, Figures~\ref{app21} and~\ref{app22} represent the second simulation, and Figures~\ref{app31} and~\ref{app32} represent the third simulation.

By comparison of these sets of figures with those presented in the main body of the paper we understand what the variance is, from one simulation to the next, that the statements presented in the paper are subject to.

There is one caveat to be mentioned in the presentation of these additional results. In the text we stated that the amount of discausality in the evolution of the ecauset is anti-correlated with the amount of time symmetry in the system. As the ecauset evolves towards time symmetry the amount of discausal moves diminishes.

It would appear that this contradicts what the panels in Figures~\ref{app12},~\ref{app22} and~\ref{app32} show, which is an increase in the length of the discausal moves (length of the red lines). This would denote an increase in the relative amount of discausal moves towards late times. However we need to take into account that, in the definition we chose, a move is discausal when it retrocedes in time by a length which is larger than the lower threshold of 1/20 of the full time range of the simulation \footnote{A lower threshold to the definition of discausality is necessary, since otherwise roughly 50\% of the jumps of ecauset would be discausal: unless a jump is purely space-like, it will always be either forward or backward in time. We are interested in the jumps which are significantly forward or backward in time compared to a standard move of the evolution at that point.}.
Therefore, by construction, there will appear to be more discausal moves as the time scale of the simulation grows. However this only means that, as the system evolves in time, there are more and larger discausal moves available with jumps backwards to the far past, and not that the relative amount of discausal moves is larger.

Also, note that some of the discausal moves picked up by the middle and last portions of evolution in each simulation, which are depicted in red in Figures~\ref{app12},~\ref{app22} and~\ref{app32} are actually moves within a limit cycle. These are regular moves and just happen to be between two elements displaced far enough in time to be classified as discausal, though they are not discausal moves in the sense of opposing the unidirectional Minkowski time evolution in which we are interested here.

\clearpage


\end{document}